\documentclass[aps, pra, a4paper, amsfonts, amssymb, amsmath, reprint, showkeys, nofootinbib, twoside, superscriptaddress, longbibliography]{revtex4-1}
\usepackage[english]{babel}
\usepackage[utf8]{inputenc}
\usepackage[colorinlistoftodos, color=green!40, prependcaption]{todonotes}
\usepackage{amsthm}
\usepackage{mathtools}
\usepackage{physics}
\usepackage{xcolor}
\usepackage{graphicx}
\usepackage{siunitx}
\usepackage[T1]{fontenc}
\usepackage{soul}

\usepackage{newtxtext,newtxmath}

\makeatletter
\newcommand{\pushright}[1]{\ifmeasuring@#1\else\omit\hfill$\displaystyle#1$\fi\ignorespaces}
\newcommand{\pushleft}[1]{\ifmeasuring@#1\else\omit$\displaystyle#1$\hfill\fi\ignorespaces}
\makeatother

\newcommand{\fig}[1]{Fig.~\ref{#1}}

\newcommand{\eq}[1]{Eq.~(\ref{#1})}

\usepackage[acronym,nonumberlist]{glossaries}
\newacronym{NP}{NP}{nanoparticle}
\newacronym{GAP}{GAP}{Gaussian approximation potential~\cite{bartok2010gaussian,klawohn_2023}}
\newacronym{SOAP}{SOAP}{smooth overlap of atomic positions~\cite{bartok2013representing}}
\newacronym{fcc}{fcc}{face-centered cubic}
\newacronym{NS}{NS}{nested sampling}
\newacronym{EAM}{EAM}{embedded-atom method}
\newacronym{DFT}{DFT}{density-functional theory}
\newacronym{ML}{ML}{machine learning}
\newacronym{MLP}{MLP}{{\gls{ML}} potential}
\newacronym{PES}{PES}{potential energy surface}
\newacronym{HER}{HER}{the hydrogen evolution reaction}
\newacronym{PBE}{PBE}{Perdew-Burke-Ernzerhof~\cite{perdew1996generalized}}
\newacronym{PBE-DFT}{PBE-DFT}{{\gls{DFT}} with the {\gls{PBE}} exchange-correlation functional}
\newacronym{MD}{MD}{molecular dynamics}
\newacronym{AIMD}{AIMD}{\textit{ab initio} {\gls{MD}}}
\newacronym{GCMC}{GCMC}{grand-canonical {\gls{MC}}}
\newacronym{HRMC}{HRMC}{hybrid {\gls{RMC}}}
\newacronym{XPS}{XPS}{X-ray photoelectron spectroscopy}
\newacronym{GW}{GW}{$GW$ theory}
\newacronym{ANN}{ANN}{artificial neural network}
\newacronym{KRR}{KRR}{kernel ridge regression}
\newacronym{MC}{MC}{Monte Carlo}
\newacronym{GPR}{GPR}{Gaussian process regression}
\newacronym[longplural={core-electron binding energies}]{CEBE}{CEBE}{core-electron binding energy}
\newacronym{ACO}{a-CO$_x$}{oxygen-rich amorphous carbon}
\newacronym{GO}{GO}{graphene oxide}
\newacronym{rGO}{rGO}{reduced graphene oxide}
\newacronym{RMC}{RMC}{reverse {\gls{MC}}}
\newacronym{XRD}{XRD}{X-ray diffraction}
\newacronym{vdW}{vdW}{van der Waals}
\newacronym{ASE}{ASE}{the Atomic Simulation Environment~\cite{larsen_2017}}
\newacronym{PAW}{PAW}{projector augmented-wave~\cite{bloechl_1994,kresse_1999}}
\newacronym{XANES}{XANES}{X-ray absorption near-edge spectroscopy}
\newacronym{SAXS}{SAXS}{small-angle X-ray scattering}
\newacronym{SI}{SI}{supporting information}
\newacronym{MBD}{MBD}{many-body dispersion~\cite{tkatchenko_2012}}
\newacronym{TS}{TS}{Tkatchenko-Scheffler~\cite{tkatchenko_2009}}
\newacronym{lMBD}{lMBD}{linear-scaling {\gls{MBD}}~\cite{muhli_2024}}
\newacronym{D2}{D2}{Grimme's D2~\cite{grimme_2006}}
\newacronym{D3}{D3}{Grimme's D3~\cite{grimme_2010}}
\newacronym{SCS}{SCS}{self-consistently screened}
\newacronym{cTS}{cTS}{corrected {\gls{TS}}}
\newacronym{RDF}{RDF}{radial distribution function}
\newacronym{XDM}{XDM}{exchange-hole dipole moment~\cite{becke_2005}}
\newacronym{DoS}{DoS}{density of states}

\begin{document}
\title{On-the-fly reparametrization of pairwise dispersion 
interactions for accurate and efficient molecular dynamics: Phase diagram of white phosphorus}

\author{Heikki Muhli}
\affiliation{Department of Chemistry and Materials Science, Aalto University, 02150 Espoo, Finland}
\affiliation{Department of Applied Physics, Aalto University, FIN-00076 Aalto, Espoo, Finland}
\author{Tapio Ala-Nissila}
\affiliation{QTF Center of Excellence, Department of Applied Physics, Aalto University, FIN-00076 Aalto, Espoo, Finland}
\affiliation{Interdisciplinary Centre for Mathematical Modelling and Department of Mathematical Sciences, Loughborough University, Loughborough, Leicestershire LE11 3TU, United Kingdom}
\author{Miguel A. Caro}
    \email{mcaroba@gmail.com}
    \affiliation{Department of Chemistry and Materials Science, Aalto University, 02150 Espoo, Finland}

\date{2 November 2024}

\begin{abstract}
\begin{center}
\textbf{Abstract}
\end{center}
Accurate estimation of the contribution from dispersion interactions to the total energy
is important for many molecular systems and low-dimensional solids. In this work we
demonstrate how the recently developed linear-scaling many-body dispersion correction
method can be efficiently applied in molecular dynamics simulations while keeping
high accuracy. This is achieved by reparametrization of the effective pairwise dispersion
interactions on the fly during the simulation. We demonstrate this method by computing
order-disorder and solid-liquid transitions of the phase diagram of white phosphorus (P$_4$).
\end{abstract}

\maketitle

\section{Introduction}

Intermolecular \gls{vdW} interactions are the glue that holds weakly bonded
systems together. For instance, graphene layers are formed by covalent bonding
but graphite would not exist if the layers were not held together by \gls{vdW} forces. Similarly, single DNA strands form the double-helix structure
because of \gls{vdW} forces. These
interactions describe the repulsion and attraction between dipoles in a chemical
system, some of which are permanent and some induced. Permanent dipoles are not
present in all molecules but induced dipoles are exhibited by all atoms
interacting with each other due to quantum-mechanical superposition of the
electronic density~\cite{truhlar2019dispersion}. The interactions between these
induced dipoles are known as the dispersion interactions, which are by far the largest contribution to the \gls{vdW} forces, and they are present in all
compounds. By magnitude, they are weak as compared to covalent bonds between atoms
but, without them, most molecules would be repelled by each other and life as
we know it would not exist.

Because these interactions arise from electron correlations, they are not
properly captured by quantum-mechanical \gls{DFT} with semi-local density functionals. In
computational atomistic modeling, varying levels of dispersion corrections can
be applied to electronic-structure calculations in post processing. Often treated
as two-body interactions between dipoles, dispersion is actually a many-body
interaction in nature. The leading pairwise term is enough for a good first
approximation of the interactions, and pairwise dispersion models are still
widely used as dispersion corrections for \gls{DFT} today because of their
computational efficiency. However, the accuracy of pairwise dispersion-correction
models is inconsistent across molecular and material systems~\cite{blood2016analytical,hermann2017first}.
As could be expected, pairwise methods, although computationally efficient,
generally perform worse than the ones that take into account the many-body effects,
such as the \gls{XDM} and \gls{MBD} methods~\cite{hermann2017first}.

Even for such small and simple systems as a benzene dimer, pairwise methods give
noticeable errors compared to \gls{MBD}~\cite{blood2016analytical,hermann2017first}, and the magnitude of
these errors is not systematic, i.e., they may depend on molecular orientation (even
for the same set of molecules). This implies that, for a proper study of a system
consisting of molecules with rotational disorder, the description of many-body effects
is required. An interesting material where rotational ordering of molecules is important is white
phosphorus, which consists of P$_4$ molecules. Phosphorus as an element has a particularly
rich phase diagram~\cite{clark2010compressibility} and structural diversity~\cite{deringer_2020b}, exhibiting graphite-like layered structures in different
allotropes of black phosphorus at increasing 
pressures~\cite{brown1965refinement,jamieson1963crystal,kikegawa1983x,akahama1999simple,akahama2000structural},
complex stacked structures of crystalline fibrous and
Hittorf's phosphorus~\cite{ruck2005fibrous,zhang2017hittorf}, amorphous red phosphorus~\cite{zhou_2023},
white phosphorus where the structure is defined by the orientation of the P$_4$
molecules~\cite{simon1987crystal,simon1997polymorphism,okudera2005crystal,clark2010compressibility},
and the molecular and polymeric liquid phases~\cite{katayama2000first}.

While dispersion interactions are important in almost any allotrope of phosphorus~\cite{deringer_2020},
their correct description is particularly important in the metastable phase diagram of white phosphorus
because of the rotational ordering of the P$_4$ tetrahedra. For a computational study of this phase
diagram with consistent accuracy, one could employ the \gls{MBD} level of dispersion corrections.
However, to properly account for disorder, a large enough supercell and sufficient configurational
sampling, e.g., with \gls{MD}, are both required. \gls{MBD}, as implemented in most \gls{DFT} codes,
scales as $\sim {\cal O}(N^3)$,
where $N$ is the number of atoms in the system. Thus, from a computational point of view, it is
infeasible to run these calculations using \gls{DFT} with \gls{MBD} corrections. On the other hand,
our recently developed \gls{lMBD} method, used in combination with the \gls{GAP} \gls{ML} framework to
avoid the \gls{DFT} calculations, is amenable to these large-scale simulations because it scales
linearly with the number of atoms in the system. However, a significant problem with \gls{lMBD} remains:
while the method is linear-scaling, the calculation of the atom-centered dispersion energy scales as
$\sim {\cal O}(N_\text{n}^2)$, where $N_{\text{n}}$ is the number of neighbors within the cut-off sphere
centered on the atom, which adds a significant overhead to the \gls{GAP} base energy calculation,
compared to using pairwise dispersion such as \gls{TS}, for example.

To circumvent this problem, we propose here a method where the \gls{lMBD}-level dispersion correction
need not be calculated at each step of an \gls{MD} trajectory. Instead, we compute the
\gls{lMBD} corrections after a given number of MD steps, say every 10 or 100 steps,
and use these data to reparametrize the pairwise
dispersion as necessary, with the aim to reproduce the more accurate \gls{lMBD} observables.
We call this method {\it on-the-fly reparametrization of the pairwise dispersion} and employ it to run
accelerated, highly efficient \gls{MD} simulations without significant loss in accuracy. We demonstrate
the applicability of the method to compute 
the metastable phase diagram of white phosphorus.

\section{Methods}

\subsection{Machine learning potentials}

\Glspl{MLP} rely on non-parametric fits to a reference \gls{PES} to enable accurate learning
and efficient prediction of energies and forces in atomistic systems, usually targeting
\gls{DFT} reference data. They are a relatively recent
development in the field of atomistic modeling~\cite{behler_2007,bartok2010gaussian}, but
have been quickly and widely adopted within the past few years because of their proven
versatility and predictive power. Excellent reviews are already available and the reader
is referred to those for further details~\cite{deringer_2019,hellstrom_2020,schutt_2020,deringer_2021}.
Here, we briefly mention the basic characteristics of the \gls{GAP} \gls{MLP} approach that
we use both to model the short-range (i.e., not accounting for \gls{vdW} interactions)
part of the \gls{PES} and to derive the local parameters of our \gls{vdW} correction models.

A local property $\xi$ of atom $i$ within an atomistic system can be expressed as a sum over kernel
functions~\cite{bartok2010gaussian,deringer_2019}:
\begin{align}
    \xi_i = \delta^2 \sum\limits_{s=1}^{N_\text{s}} \alpha_s k(s,i),
    \label{13}
\end{align}
where $k(s,i)$ is the kernel, giving a measure of similarity between the environments of
atoms $i$ and $s$, $\alpha_s$ is a fitting coefficient, $\delta$ defines the scale of the
local property, and $s$ runs over $N_\text{s}$ ``sparse'' atomic configurations in the
training set. The kernel functions are often computed within the \gls{GAP} formalism
from the \gls{SOAP} descriptor. A central assumption in \gls{GAP} (and other \gls{MLP}
approaches) is that the total potential energy of the system $E$ can be decomposed in terms
of local energies $\epsilon_i$ computed as in \eq{13}:
\begin{align}
    E = \delta^2 \sum\limits_{i=1}^N \epsilon_i,
\end{align}
where $N$ is the total number of atoms in the system. This is an uncontrollable, although
intuitive, approximation, as the total energy is a physical observable but the local energies
do not have a physical meaning. They are, mathematically, given by the optimal atom-wise
decomposition
of the total energies in the training set according to some objective function defined during
the training stage. The approximation of locality can be tested empirically~\cite{deringer_2017}
and is otherwise widely considered to work reliably for covalent interactions.

In this paper, we use previously developed~\cite{muhli2021machine,muhli_2024} models trained
with the GAP code~\cite{klawohn_2023}, and run the calculations with the TurboGAP atomistic
simulation engine~\cite{caro_2019,ref_turbogap}. We recomputed the phosphorus database from
Ref.~\cite{deringer_2020} with the \gls{PBE} functional and the VASP
code~\cite{kresse1996efficient}, and refitted a phosphorus \gls{GAP} from it. The Hirshfeld
volumes obtained from that database were used to train a local-property model to enable the
addition of \gls{vdW} corrections~\cite{muhli2021machine}.
The new \gls{MLP} has been made freely available online~\cite{muhli_2024b}.
In addition, we use ASE~\cite{larsen_2017},
Ovito~\cite{stukowski_2009} and various in-house codes for structural manipulation, analysis
and visualization.

\subsection{Linear-scaling pairwise and many-body vdW corrections}

With semi-local exchange-correlation functionals used in \gls{DFT},
the long-range correlation is not captured properly~\cite{martin2008electronic,stohr2019theory}.
Most of the long-range correlation energy is attributed to dispersion, and there exist various
ways to include it in a \gls{DFT} calculation, from non-local
functionals~\cite{dion_2004,klimes_2010} to dispersion corrections applied in
post-processing~\cite{grimme_2006,grimme_2010,tkatchenko_2009,tkatchenko_2012,becke_2005,otero_2013}.
Non-local functionals are often expensive and have historically involved averaging the density (ADA~\cite{gunnarsson1976exchange}) and weighting the electronic density (WDA~\cite{gunnarsson1977exchange,alonso1978nonlocal,gunnarsson1979descriptions}) and thus have serious problems with core electrons distorting the weighting or averaging in an unphysical way.
However, there have been some promising results for non-local functionals more recently, for example in
Refs.~\onlinecite{dion_2004,roman2009efficient,klimevs2011van,hermann2020density}, but they are still computationally expensive as compared to semi-local functionals. The corrections added in post-processing have been more popular
among the \gls{DFT} community because of their computational efficiency and the fact that they
are optimized for widely used exchange-correlation functionals such as \gls{PBE}, making them
easy to use. They also allow for easier local parametrization of the properties of the system,
making them ideal for \gls{MLP}s that usually rely on local descriptors of the atomic
environments~\cite{bereau_2018,muhli2021machine,ying_2023,tu_2024,muhli_2024}.

The most common type of dispersion correction applied in post-processing is the pairwise dispersion,
due to its computational efficiency and the fact that it is the leading term in the dispersion
interactions. In the pairwise model, the interatomic dispersion interaction between a pair of atoms
has the widely recognized $1/r_{ij}^6$ dependence, where $r_{ij}$ is the interatomic distance
between atoms $i$ and $j$. The strength of this interaction is determined by the atomic
polarizabilities of atoms $i$ and $j$, and these polarizabilities can be calculated in various
ways. An example of a pairwise dispersion correction used with \gls{DFT} is the \gls{TS} method,
where the dispersion energy is given by
\begin{equation}
    E_{\text{TS}} = -\frac{1}{2}\sum\limits_{i=1}^N\sum\limits_{j\neq i}
    \frac{C_{6,ij}}{r_{ij}^6} f_{\text{damp}}(r_{ij};R_{\text{vdW},ij},d,s_{\text{R}}),
\end{equation}
where the sum goes over all $N$ atoms in the system, $f_{\text{damp}}$ is a damping function used
to suppress the interactions at short range that are already captured by the semi-local
exchange-correlation functional, $d$ and $s_{\text{R}}$ are parameters fitted for the functional used,
and $R_{\text{vdW},ij}$ and $C_{6,ij}$ are coefficients that are calculated from free atomic
values and effective atomic volumes obtained from partitioning of the electronic
density~\cite{hirshfeld1977bonded,tkatchenko_2009}.
These effective volumes can be considered as local parameters of the system and we have previously
shown that they can be predicted with \gls{ML}, using a local-property model~\cite{klawohn_2023}
amenable to the \gls{GAP} framework, with many-body \gls{SOAP} descriptors.
This allows us to apply the \gls{TS} corrections on top of a \gls{MLP} for accelerated
\gls{MD} simulations with \gls{vdW} corrections~\cite{muhli2021machine}. The pairwise model is a
good approximation for a wide variety of systems but fails to capture some physics that arise
from proper many-body effects~\cite{blood2016analytical,hermann2017first}.

For this reason, we have recently extended the model to include \gls{MBD}
interactions~\cite{muhli_2024}. The model is physics-based such that it only takes as
arguments the effective atomic volumes, called effective Hirshfeld volumes. These effective
volumes can be calculated with \gls{DFT} or predicted with \gls{GAP} (or another atomistic \gls{ML}
flavor). The original \gls{MBD} method was formulated from the coupled fluctuating dipole
model~\cite{tkatchenko_2012} and later reformulated from the
adiabatic-connection-fluctuation-dissipation theorem and random-phase
approximation~\cite{tkatchenko2013interatomic}. The reformulation~\cite{buvcko2016many} is also
used in VASP~\cite{kresse1993ab,kresse1996efficiency,kresse1996efficient}, the most widely
used \gls{DFT} code,
which we also use as our reference method. The \gls{MBD} dispersion energy in this formulation
is given by
\begin{equation}
    E_{\text{MBD}} = \int\limits_{0}^\infty \frac{\mathrm{d}\omega}{2\pi}
    \mathrm{Tr}\left\{ \ln \left( \mathbf{1} - \mathbf{A}_{\text{LR}}(\omega)
    \mathbf{T}_{\text{LR}} \right)\right\},
    \label{eq:E_MBD}
\end{equation}
where $\mathbf{A}_{\text{LR}}(\omega)$ is a diagonal frequency-dependent matrix that contains
the polarizabilities, which are solved for self-consistently in this case but still involve
effective Hirshfeld volumes. $\mathbf{T}_{\text{LR}}$ is the dipole-coupling tensor. Effectively,
the logarithm in Eq.~\eqref{eq:E_MBD} includes many-body interactions between the atoms in the
system up to infinite body order, which becomes apparent if one expands the logarithm as a series.

The truncation of this series serves as the starting point for our atom-centered formulation of
\gls{MBD} which we call \gls{lMBD}. The atom-centered picture, instead of the ``entire system''
approach that the reference method takes, is necessary to enable computational linear scaling
with respect to the number of atoms in the system, a prerequisite to make these simulations
tractable for large systems. In \gls{lMBD}, the \emph{local} \gls{vdW} energy of atom $k$ is given by
\begin{equation}
    \epsilon_{K,k} = -\int\limits_0^\infty \frac{\text{d}\omega}{2\pi} \sum\limits_{n=2}^{n_{\text{max}}} c_n \mathrm{Tr}\left\{\mathbf{g}_{K,k}\mathbf{G}_K^{n-2}\mathbf{g}_{K,k}^T\right\},
\end{equation}
where the sum arises from the series expansion of the logarithm in Eq.~\eqref{eq:E_MBD},
truncated at some maximum body order. We use the capital letter $K$ to denote the atomic
environments (and sets of indices) where we are operating: atom $k$ is the central atom of
atomic environment $K$. Here $\mathbf{G}_K \in \mathbb{R}^{3 N_{\text{n}} \times 3 N_{\text{n}}}$
is the symmetrized $k$-centered version of $\mathbf{A}_{\text{LR}}(\omega)\mathbf{T}_{\text{LR}}$
that only includes elements (number of neighbors $N_{\text{n}}$) up to some cut-off radius from
the central atom. $\mathbf{g}_{K,k} \in \mathbb{R}^{3 \times 3 N_{\text{n}}}$ contains only the rows
corresponding to the central atom $k$. The coefficients $c_n$ are either derived from the logarithm
series, where they are given as $c_n = 1/n$, or they can be fitted to the range of eigenvalues
as we show in Ref.~\onlinecite{muhli_2024}. The total dispersion energy is given as a sum of the local
energy contributions from each central atom:
\begin{equation}
    E_{\text{lMBD}} = \sum\limits_{k=1}^N \epsilon_{K,k},
\end{equation}
where the summation is done such that $k$ is always the central atom and $\epsilon_{K,k}$ thus
comes from the $k$-centric contribution. In addition to the energies, we also define the \gls{SCS}
polarizabilities (used in $\mathbf{A}_{\text{LR}}(\omega)$ in Eq.~\eqref{eq:E_MBD}) in atom-centered
fashion to make the whole calculation linear-scaling. We will omit the details here but they can be
found in Ref.~\onlinecite{muhli_2024}.

\subsection{Pairwise forces and the virial within the lMBD formalism}

For the purpose of running \gls{MD} simulations, forces are required. In addition, to estimate
the pressure for the purpose of running an $NPT$ simulation with an \gls{MD} ``barostat'', or to
simply monitor its instantaneous value during an $NVT$ run, it is computationally advantageous
to compute the virial tensor.

As usual, the forces are given as the negative
gradient of the energy with respect to the atomic positions.
We calculate them by taking the derivative of the total energy in atomic
environment $K$ with respect to the position of the central atom. There are certain intricacies related
to the derivatives which we omit here. In essence,
we make an approximation that neglects the gradients of the polarizabilities if they are not directly
coupled to the central atom to bring the computational cost of the force calculation to the same level
with the local-energy calculation. Furthermore, we treat the $n=2$ term separately, because it is
cheaper to calculate with a Hadamard product (element-wise product), which we denote by ``$\odot$''.
For details, we refer the reader to Ref.~\onlinecite{muhli_2024} and we proceed to give the expression for
the \gls{lMBD} forces here:
\begin{multline}
    f_{k}^\gamma \approx -\int\limits_0^\infty \frac{\mathrm{d}\omega}{\pi} \Bigg[c_2 \sum\limits_{i,j}\sum\limits_{\alpha,\beta}\left[\mathbf{G}_K \odot \frac{\partial\mathbf{G}_K}{\partial r_k^\gamma}\right]_{ij}^{\alpha\beta} \\+\sum\limits_{n=3}^{n_{\text{max}}} c_n n \mathrm{Tr} \left(\mathbf{g}_{K,k}\mathbf{G}_K^{n-2} \frac{\partial \mathbf{g}_{K,k}^T}{\partial r_k^\gamma}\right)\Bigg].
    \label{eq:cent-appr}
\end{multline}

We derive the instantaneous atomic pressures from the average of the sum of the diagonal elements
of the zero-temperature stress tensor $\boldsymbol{\sigma}$:
\begin{align}
    P = & - \frac{1}{3} \mathrm{Tr}\{\boldsymbol{\sigma}\} + \frac{N k_\text{B} T}{V}; \nonumber \\
    \boldsymbol{\sigma} = & - \frac{1}{V} \boldsymbol{\phi},
    \qquad \boldsymbol{\phi}, \boldsymbol{\sigma} \in \mathbb{R}^{3\times 3},
\end{align}
where $P$, $T$ and $V$ are the (instantaneous) pressure, temperature and volume, respectively, and
$k_\text{B}$ is Boltzmann's constant.
The literature is somewhat inconsistent in terms of how the definition of the virial (stress) tensor
is done, and whether the thermal component is included or not. In our definition above we have chosen
to separate the contributions to the pressure
arising from the potential energy and those arising from the thermal kinetic energy, and to notate the
virial tensor $\boldsymbol{\phi}$, with units of energy (e.g., eV/atom).
The calculation of the virial tensor for many-body interatomic potentials is not, in general,
a trivial problem. A couple of derivations are given, for instance, in Refs.~\onlinecite{thompson2009general}
and \onlinecite{fan2015force}. Fortunately, some functional forms admit a useful simplification:
if one is able to express the forces using pairwise contributions, the virial tensor is then given
by the same equation that is valid for a general pairwise (i.e., two-body)
potential~\cite{thompson2009general,subramaniyan2008continuum}:
\begin{equation}
    \phi^{\alpha\beta} = - \frac{1}{2} \sum_{i=1}^N \sum\limits_{j \neq i}^N(r_j^\alpha - r_i^\alpha) f^\beta_{ij},
    \label{eq:virial}
\end{equation}
where $\alpha$ and $\beta$ denote the Cartesian components,
$r_i^\alpha$ is the Cartesian component $\alpha$ of the position of atom $i$, and $f_{ij}^\beta$
is the Cartesian component $\beta$ of the pairwise force on atom $i$ due to
atom $j$. To estimate the \gls{lMBD} contribution to the virial tensor we thus only need
to define the pairwise forces for \gls{lMBD}.

Because we already partitioned the dispersion energy into atom-wise contributions, this is quite easy: we take the contribution of atom $i$ to the total energy in the force calculation. That is:
\begin{equation}
    f_{ik}^\gamma = -\frac{\partial \epsilon_{K,i}}{\partial r_k^\gamma}.
\end{equation}
What this looks like in our expression for the forces in Eq.~\eqref{eq:cent-appr} with the same approximations applied is:
\begin{multline}
    f_{ik}^\gamma \approx -\int\limits_0^\infty \frac{\mathrm{d}\omega}{\pi} \Bigg[c_2 \sum\limits_{j}\sum\limits_{\alpha,\beta}\left[\mathbf{G}_K \odot \frac{\partial\mathbf{G}_K}{\partial r_k^\gamma}\right]_{ij}^{\alpha\beta} \\+\sum\limits_{n=3}^{n_{\text{max}}} c_n n \mathrm{Tr} \left(\left[\mathbf{g}_{K,k}\mathbf{G}_K^{n-2}\right]_i \frac{\partial \mathbf{g}_{K,ki}^T}{\partial r_k^\gamma}\right)\Bigg].
\end{multline}
Here we only take the rows of $\mathbf{g}_{K,k} \mathbf{G}^{n-2}_K$ corresponding to atom $i$: $[\mathbf{g}_{K,k} \mathbf{G}^{n-2}_K]_i \in \mathbb{R}^{3\times 3}$ and multiply with the corresponding derivative elements of $[\mathbf{g}_{K,k}]_i = \mathbf{g}_{K,ki} \in \mathbb{R}^{3\times 3}$. For the $n=2$ term, we only take the sum over elements corresponding to atom $i$.

\subsection{On-the-fly reparametrization of pairwise vdW corrections}\label{11}

\begin{figure}
\centering
\includegraphics[width=\linewidth]{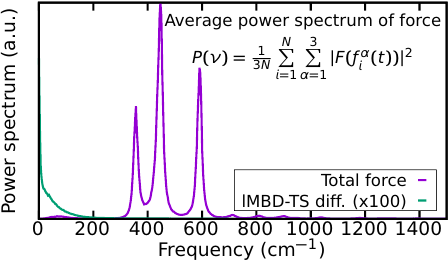}
\caption{Average power spectrum computed from the Fourier transform of the atomic
force components in a liquid white phosphorus
system with 125 P$_4$ molecules. The frequency decomposition is obtained from
a 10~ps trajectory with 2~fs sampling. The ``total force'' shows the curves derived
from GAP+TS forces whereas the ``lMBD-TS'' difference shows a magnified ($\times 100$
higher) representation of the curves derived by Fourier transforming the difference
in TS and MBD forces.}
\label{04}
\end{figure}

Even within the computationally advantageous \gls{lMBD} approximation, the cost of
\gls{vdW} corrections is significant, especially when compared to other \gls{vdW}
correction schemes based on effective pairwise interactions like \gls{TS},
\gls{D2} and \gls{D3}. For instance, we have noticed a slowdown of between one and
two orders of magnitude in our TurboGAP implementation of \gls{lMBD}~\cite{muhli_2024} compared to
a TurboGAP \gls{TS} calculation~\cite{muhli2021machine} with typical \gls{lMBD} parameters. The slowdown
is strongly affected by atom density and cutoffs, which together determine the
typical number of neighbor atoms $N_\text{n}$ within the \gls{lMBD} cutoff
sphere, with the method scaling as $\mathcal{O}(N {N_\text{n}}^2)$
(\gls{TS} scales as $\mathcal{O}(N N_\text{n})$, for comparison).
To tackle this issue of computational speed, we take advantage of the fact that the
\textit{differences} between \gls{vdW} correction schemes evolve more slowly than
other properties of the system. For instance, \fig{04} shows the average power spectrum
(the squared modulus of the Fourier transform) of the atomic forces in a liquid
white phosphorus test case. The spectrum shows the main vibrational peaks where they
are expected based on existing literature~\cite{gutowsky_1950,chikvaidze_2023} for the
total force,
and a very weak ($\times 100$ magnification in the figure) diffusive component at low frequency
for the \gls{vdW} force.
This clearly indicates that the difference in \gls{lMBD} and \gls{TS} forces evolves
within significantly longer time scales than the vibrational dynamics, and thus motivates
the possibility to apply a \gls{vdW} correction scheme that gets updated less frequently than
the remaining part of the potential energy, i.e., a multistepping approach~\cite{streett_1978}.

On the one hand, the multistep correction to the total \gls{vdW} energy $E_\text{vdW}$
and the ``total'' (in the sense of averaged over the atoms in the simulation cell) \gls{vdW} pressure
$P_\text{vdW}$ can be implemented trivially:
\begin{align}
E_\text{cTS} (t_k) = E_\text{TS} (t_k) + (E_\text{MBD} (t_{k'}) - E_\text{TS} (t_{k'}));
\label{05}
\\
P_\text{cTS} (t_k) = P_\text{TS} (t_k) + (P_\text{MBD} (t_{k'}) - P_\text{TS} (t_{k'}));
\label{06}
\\
\quad \text{where} \quad k' = k - (k \bmod N_\text{c}). \nonumber
\end{align}
Here, $N_\text{c}$ is the number of \gls{MD} time steps in between corrections, i.e.,
a \gls{TS} calculation is carried out at every time step, $k$, and a \gls{MBD}
calculation is carried out only every $N_\text{c}$ time steps, and the correction
is fixed as a constant shift in either the \gls{vdW} energy or pressure during the
following $N_\text{c}$ time steps. The justification in both cases is that these
corrections are small, compared to the overall oscillations in the potential energy
and pressure, and change slowly, as we will see in the next section
with some examples. We also note that, in practice, we do not compute the pressure
corrections but the corrections to the full stress tensor $\boldsymbol{\sigma}$. We
refer to this approach as \gls{cTS}.

Another important point to note is that we apply these corrections for the purpose
of: (i) in the case of the energy, estimating an ``\gls{MBD}-quality'' approximation
to the \gls{vdW} energy, useful for energy comparison across different simulations;
and (ii) in the case of the pressure, to enable correct barostatting during \gls{MD},
as the contribution of \gls{vdW}
corrections to the pressure (and thereby the density)
can be rather significant within the low-pressure regime, especially for molecular
systems. The \gls{vdW} forces,
on the other hand, determine the detailed dynamics of the system, via solving
Newton's equations of motion. Applying a naive correction to the forces in the
spirit of Eqs.~(\ref{05}) and (\ref{06}) does not work for two reasons. First, the
\gls{vdW} forces are computed individually for each atom and the averaging
effect that stabilizes the corrections to $E_\text{vdW}$ and $P_\text{vdW}$ is
lost. Second, the \gls{MBD} and \gls{TS} methods have very different short-range
behavior, in part due to their different damping function parametrization, with
the interaction with nearest-neighbor atoms driving rather noisy \gls{vdW}
force differences from which a correction cannot be as effectively applied as for
the energy and pressure. In this work, instead of working directly with the forces
and their time derivatives, as in established multistepping approaches~\cite{streett_1978},
we proceed according to what we call an ``on-the-fly reparametrization'' approach.

To retrieve the approximate \gls{MD} behavior without resorting to direct
approximation of the forces, we take advantage of the fact that the virial
tensor has a straightforward local (atom-wise) decomposition and that, in the virial,
pairwise contributions are given by the pairwise forces \textit{weighted} by the
pairwise distance. This allows us to define a damped version of the local virial
tensor with the same smooth cutoff function defined for both \gls{TS} and \gls{MBD}:
\begin{align}
\tilde{\phi}_i^{\alpha \beta} = - \frac{1}{4} \sum_j \left( f_{ij}^\alpha r_{ij}^\beta +
f_{ij}^\beta r_{ij}^\alpha \right) f_\text{cut}(r_{ij}; r_\text{min}, r_\text{max}),
\label{08}
\end{align}
where the factor of 1/4 accounts for both double counting and symmetrization,
$r_{ij}^\alpha := r_j^\alpha - r_i^\alpha$, $r_{ij}$ is the scalar distance between
atoms $i$ and $j$, and
\begin{align}
& f_\text{cut}(r_{ij}; r_\text{min}, r_\text{max}) =
\nonumber
\\
&
\begin{cases}
0 \quad \text{if} \quad r_{ij} < r_\text{min};\\
1 \quad \text{if} \quad r_{ij} > r_\text{max};\\
\frac{1}{2} + \frac{3}{4} \frac{ 2r_{ij} - \left( r_\text{max}
+ r_\text{min}\right)}{r_\text{max}-r_\text{min}}
- \frac{1}{4} \frac{\left( 2 r_{ij} - \left( r_\text{max} + r_\text{min} \right) \right)^3}
{\left( r_\text{max} - r_\text{min} \right)^3} \quad \text{else}.
\end{cases}
\label{09}
\end{align}
The parameters $r_\text{min}$ and $r_\text{max}$ determine the region where the
damping function goes from no coupling ($f=0$) to full coupling ($f=1$).
The tilde in $\tilde{\boldsymbol{\phi}}_i$ indicates that this is the damped version
of the local virial, and not the full local virial. Then, the correction to
the \gls{TS} interactions is derived in terms of finding some
\gls{TS} parametrization that approximates
\begin{align}
\frac{\tilde{\phi}_{i,\text{TS}}^{\alpha \beta} (\{ \tilde{C}_{6,i} \})}
{\tilde{\phi}_{i,\text{MBD}}^{\alpha \beta}} \approx 1,
\end{align}
where $\tilde{C}_{6,i} := s_i C_{6,i}$ is the scaled London-dispersion coefficient
for atom $i$, with $s_i$ being the scaling factor. The scaling factors are recomputed every
$N_\text{c}$ steps and the \gls{vdW} forces are obtained from the reparametrized
\gls{cTS} model. After exploring different ways to estimate the $\{s_i\}$, we
settled for the following heuristics:
\begin{align}
& s_i (t_{k}) = \text{clip}\bigg( s_i (t_{k' -1}) \nonumber
\\
& \qquad + 0.1 \times \Big(
\text{clip}\Big(\frac{\text{Tr}(\tilde{\phi}_{i,\text{MBD}}(t_{k'}))}{\text{Tr}(\tilde{\phi}_{i,\text{TS}}(t_{k'}))}; -1, 3 \Big) - 1 \Big); 0.1, 2.5 \bigg),
\label{07}
\end{align}
where the initial value of the scaling factor is $s_i(t_{-1}) = 1$, $k' = k-(k\bmod N_{\text{c}})$ and the $\text{clip}(x;x_1,x_2)$ function returns $x$ clipped onto the $[x_1,x_2]$
interval, i.e., if $x$ is within the interval it returns $x$, otherwise it returns
whichever value $x_1,x_2$ is closer to $x$.
While seemingly complicated, the idea is simple. The scaling factors are updated
slowly during the \gls{MD} run, such that the scaling factor at time $t_k$
is obtained from the scaling factor at time $ t_{k'-1} $ by increasing it
or decreasing it by an amount that improves the instantaneous agreement between
the traces of the local \gls{MBD} and \gls{TS} damped virials. To stabilize
the procedure, the actual ratio of the traces is constrained to the $[-1,3]$
interval and the ``mixing'' of the previous and currently optimal scaling factors
is restricted to 10\% per update. Finally, we impose the physical constraint
that $\tilde{C}_{6,i}$ must be bound between 0.1 and 2.5 times the value of the
uncorrected $C_{6,i}$. This somewhat convoluted procedure is needed because
instantaneously the ratio of the virials can take very large values (e.g., when
the denominator approaches zero) or have a negative sign. In the next section
we show an example of how this approach allows to correct the qualitatively
wrong behavior of \gls{TS} in white phosphorus at low density and moderate
temperature, where \gls{TS} predicts a gaseous phase whereas \gls{MBD}
predicts clustering of P$_4$ molecules.

The proposed scaling procedure only rescales the $C_6$ coefficients, not the
effective Hirshfeld volumes used to parametrize (in addition to the $C_6$
coefficients) other quantities used by \gls{TS}. We opt for this approach since
the virial tensor is approximately linear in the $\{ C_{6,i} \}$ but scaling of
the \gls{vdW} radii, which are also affected by the Hirsfeld volumes in \gls{TS},
would introduce non-trivial nonlinearities via the dependence of the
damping function on the \gls{vdW} radii. In our tests, we observed that
rescaling coupled with manual tuning of the damping function outperforms the
proposed approach. However, this strategy is non-systematic: e.g.,
a system with the same chemical makeup might require different optimal
damping function parameters at different thermodynamic conditions.
For these reasons, we are currently developing a more transferable
(e.g., across chemical species), but conceptually very similar, approach based on the
on-the-fly self-consistent reparametrization of \gls{TS}. The objective
of this generalized model is to avoid the heuristics involved in
manually defining the different parameters needed in Eqs.~(\ref{08}),
(\ref{09}) and (\ref{07}) as well
as manual tuning of the damping function. These
developments will be reported elsewhere.

Finally, we note that time-dependent Hamiltonians like the one derived from
the approach proposed here are strictly non-conservative. This means that
\gls{MD} within the \gls{cTS} method will result in energy drift regardless
of whether the $NVE$ ensemble is chosen or the choice of integrator
(e.g., velocity Verlet here). This drift will be usually small as the \gls{vdW}
forces are much smaller than the total forces. However, the user must keep
this issue in mind and carry out appropriate testing if energy drift is
expected to become an issue, e.g., for simulations at very low temperatures.

\section{Benchmarks}

\begin{figure}[t]
\centering
\includegraphics[width=\linewidth]{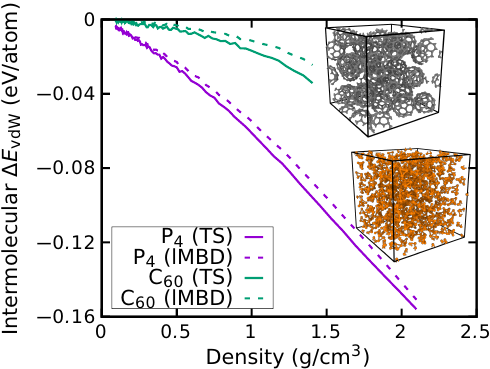}
\caption{Van der Waals contribution to the intermolecular binding energy computed as the 
difference between the total vdW energy and the sum of the vdW energies of the individual molecules.
The simulations were carried out by quickly (200~ps) compressing two hot (1000~K) molecular gases
made of 27 C$_{60}$ molecules, on the one hand, and 405 P$_4$ molecules, on the other.}
\label{01}
\end{figure}

To benchmark our methodology we have chosen white phosphorus for several reasons. First, the
strength of \gls{vdW} interactions in phosphorus is particularly large, e.g., as compared to carbon, another chemical element for which \gls{vdW} interactions are important. Phosphorus atoms are relatively large, and their valence electrons are further from the nucleus on average and held less tightly, easily forming induced dipoles due to other atoms in the vicinity. This tendency to form induced dipoles is known as the polarizability of the atom, which determines the strength of the dispersion interactions. The experimentally measured free-atomic polarizability of phosphorus is 3.630~\AA$^{3}$ and the polarizability of a carbon atom is 1.760~\AA$^{3}$~\cite{miller1978atomic}, which means that P forms dipoles more easily.
Compare for instance the simulated \gls{vdW} energy per atom in a white phosphorus gas under
rapid compression to that of C$_{60}$, shown in \fig{01}. At the density of the condensed phases,
$\approx 1.6$~g/cm$^3$ for C$_{60}$ and $\approx 1.8$~g/cm$^3$ for P$_{4}$, the average
intermolecular binding energy due to the \gls{vdW} correction is about three times larger for
P$_{4}$. This is also partially explained by the shape and size of the molecules: C$_{60}$ is a large spherical molecule with a relatively flat surface as compared to the small P$_{4}$ tetrahedron that has a lower symmetry of the wave functions.
The polarizability of a C$_{60}$ molecule is reported to be $\approx 79$~\AA$^{3}$ according to experiments~\cite{ballard2000absolute} while the polarizability of P$_{4}$ is experimentally reported as 13.589~\AA$^{3}$~\cite{thakkar2015well}. This means that the difference in polarizability \emph{per atom} is even higher when the atoms are part of these molecules, indicating that the structure of C$_{60}$ stabilizes the polarizability better than that of P$_{4}$. The second reason for choosing phosphorus for our benchmarks is that,
being an elemental, and thus homopolar, compound, the \gls{MBD} formalism with
``regular'' (as opposed to iterative~\cite{bucko_2013} or fractionally ionic~\cite{gould_2016}) effective Hirshfeld volume partitioning is sufficient to describe atomic polarizabilities
in white phosphorus. Third, the relatively large atomic mass of phosphorus means that, unlike in
hydrocarbons, quantum nuclear effects can be disregarded during the \gls{MD} simulations, thus
simplifying the configurational sampling. Fourth and last, elemental phosphorus displays a rich
phase diagram~\cite{clark2010compressibility,simon1987crystal,simon1997polymorphism,okudera2005crystal,katayama2000first}, including metastable configurations, where \gls{vdW}
interactions have a prominent role.
Thus, elucidating the atomic-scale processes that take place during
phase transitions and structural (e.g., order-disorder) transformations in white phosphorus is of
fundamental interest and opens the door to tackling the more complex structural landscape of
elemental phosphorus across all its many allotropes within a unified simulation framework.

\subsection{VdW pressure differences in the condensed phase}

\begin{figure}[t]
\centering
\includegraphics[width=\linewidth]{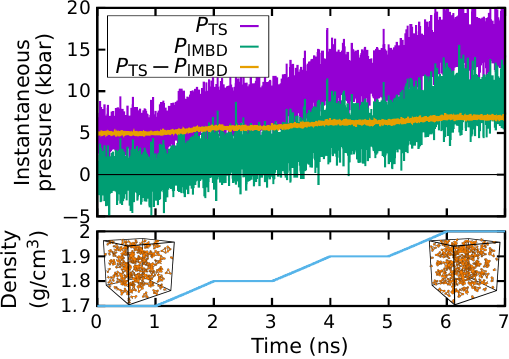}
\caption{(Top) instantaneous pressure computed for liquid white phosphorus at $T = 500$~K
as its density is increased from 1.7 to 2 g/cm$^3$ in steps (shown in the bottom panel). The
pressure is estimated, on the same atomistic structures consisting of 125 P$_4$ molecules,
following the TS and lMBD approaches. Also the difference between TS and lMBD pressures is
monitored.}
\label{02}
\end{figure}

The main justification for using a multistepping approach to account for \gls{vdW}
corrections while barostatting during \gls{MD} simulations
is provided through the example in \fig{02}. There, a 500-atom white phosphorus
system (125 P$_4$ molecules) is kept at $\approx 500$~K while its density is increased
in steps through box rescaling: we start (after preequilibration) at 1.7~g/cm$^3$
and hold it there for 1~ns, then slowly increase to 1.8~g/cm$^3$ over 1~ns, hold it
there for another 1~ns, and so on until the final density of 2~g/cm$^3$ is reached.
Throughout this process, where the interatomic energies and forces are computed within
the GAP+TS approach~\cite{muhli2021machine},
the pressures predicted by the \gls{TS} and \gls{lMBD} approaches
are monitored. For consistency, and to enable direct comparison, the pressures are
computed on the same structures. There is a large difference between the \gls{TS}-
and \gls{lMBD}-predicted values, increasing from $\Delta P \approx 5$~kbar at 1.7~g/cm$^3$
to $\Delta P \approx 7$~kbar at 2~g/cm$^3$. However, this difference is rather stable: while
the individual instantaneous pressures derived from the virial stress tensor fluctuate
very strongly, $\langle P - \langle P \rangle \rangle \approx 1.6$~kbar, the \textit{difference}
between $P_\text{TS}$ and $P_\text{lMBD}$ is comparatively rather stable,
$\langle \Delta P - \langle \Delta P \rangle \rangle \approx 0.14$~kbar. This means that
an \gls{MD} simulation at the GAP+TS level where the pressure is corrected at every step
with the \gls{lMBD} value and the value of the correction is updated every certain
number of steps (e.g., every 100 steps with a 2~fs time step) would provide rather
accurate pressure estimates at very modest computational cost. Here, it is crucial to
remark that this argument is valid for \textit{condensed} (or dense) systems, where the
fine details of the atomic trajectories are not important in terms of obtaining the statistical
properties of the system. More precisely, we refer to the role of \gls{lMBD} vs \gls{TS}
forces on the results, which is small in liquids and solids because the contribution
to the total atomic forces arising from the \gls{vdW} correction is very small as compared to
the total magnitude of the forces. For instance, in the 1.7~g/cm$^3$ example from \fig{02},
the fluctuations in the total atomic forces are
$\langle f - \langle f \rangle \rangle \approx 0.7$~eV/\AA{},
whereas the fluctuations in the difference between \gls{TS} and \gls{lMBD} forces are
$\langle \Delta f_\text{vdW} - \langle \Delta f_\text{vdW} \rangle \rangle \approx 0.03$~eV/\AA{}.

Thus, in condensed phases at $T \gg 0$~K, a \gls{TS} simulation with multistepping
correction based on the difference between \gls{TS} and \gls{MBD} pressures,
for the purpose of barostatting, should be enough to drive the system towards the same density
as predicted in an \gls{MBD}-based simulation. This assertion implicitly assumes that
the \textit{local} pressure and/or density fluctuations are relatively small and thus
do not drive the dynamics. In the dilute (gaseous) case,
this is not necessarily true, in particular if there is molecular clustering/adsorption taking place,
which we address in the next section. An additional circumstance where the details of the
dynamics may matter is whenever rotational molecular order is important in determining the
atomistic structure of the system. This could be relevant at, e.g., low-temperature conditions,
even for the solid and liquid phases.

\subsection{Molecular clustering in the dilute limit}

In the dilute (i.e., very low-density or gaseous) limit, molecular clustering can occur,
driven by \gls{vdW} interactions. In this case, the clustering behavior is influenced by
both the overall pressure \textit{and} the details of the intermolecular interactions,
which can be understood in terms of an effective ``local'' pressure.
If external constraints (e.g., container walls) are removed, a system under positive
pressure will tend to expand whereas a system under negative pressure will tend to
contract. Thus, we
can understand the tendency of molecules to agglomerate and form clusters as a response to
an effective negative local pressure, which induces local compression, and
the tendency of molecular clusters to
dissociate can be seen as a response to an effective positive local pressure, which induces
local expansion. Within our methodology, it is possible to define a local virial
tensor and, in turn, this enables a direct connection to the concept of local pressure.

We illustrate the issue of how to
correctly reproduce clustering/condensation with our white phosphorus
example. We found that, at low density and $\approx 550$~K, \gls{lMBD}
favors clustering of P$_4$, whereas this behavior is not seen within
the \gls{TS} approximation.
Note that this is indeed intuitive, as we saw that \gls{TS}
tends to predict higher pressures than \gls{lMBD} (\fig{02}), and we thus expect that
P$_4$ clusters will have a higher tendency to dissociate when modeled with \gls{TS}
corrections, as the local pressure is more positive than with \gls{lMBD}. Therefore, we
examine how this qualitatively different behavior seen at low density
can be retrieved with the on-the-fly reparametrization.

\begin{figure}
\centering
\includegraphics[width=\linewidth]{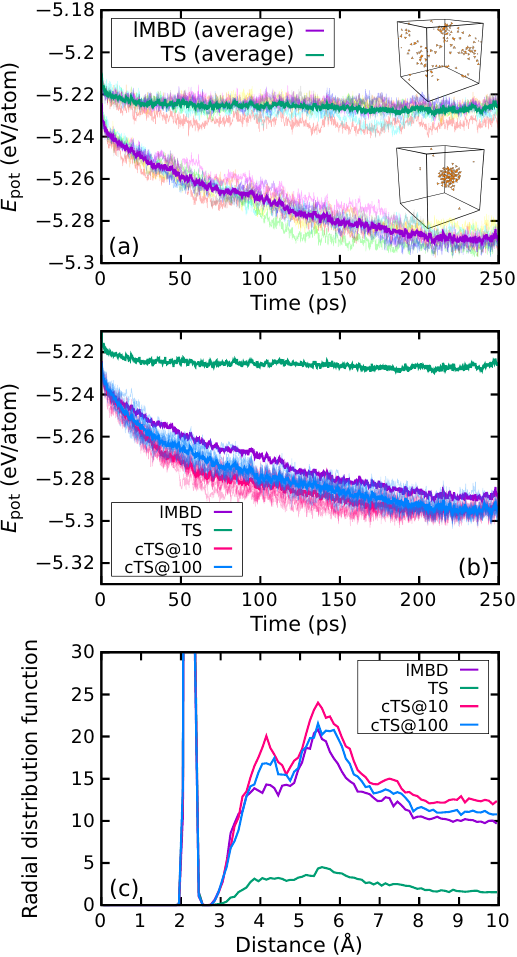}
\caption{(a) Potential energy of a dilute (gaseous) white phosphorus system
(125 P$_4$ molecules) that is initially at 800~K and then rapidly quenched to
550~K and let to evolve at that temperature for $\approx 250$~ps. The thin lines
show the potential energy evolution for 10 independent runs and the thick
lines the average over runs. The calculations are done separately for TS and
lMBD. The insets show the typical final structures for the TS runs (the system
remains gaseous) and the lMBD runs (the system aggregates into clusters). (b)
Same as in (a) but now adding results obtained with the cTS method with two
different reparametrization frequencies: 10 and 100 MD steps, corresponding to
20 and 200~fs in this case. The TS and lMBD averages from (a) are also shown
for reference. (c) Average radial distribution functions at the end of the MD
runs from (a) and (b). The first large peak corresponds to the intramolecular
distances in the P$_4$ molecules and is virtually the same for all methods. The
plateaus starting at $\approx 3$~\AA{} denote the distribution of
intracluster/intermolecular distances.}
\label{03}
\end{figure}

First, we thermalize an ensemble of 125 P$_4$ molecules at 800~K, well above the
temperature at which clustering is observed. We set the density to
$\approx 0.38$~g/cm$^3$, low enough that the system will remain in the gaseous state
but high enough to increase the frequency of molecular encounters so that the
clustering process takes place within reasonable \gls{MD} times. Ten such
uncorrelated configurations
are generated by sampling individual snapshots at 200~ps intervals from the same trajectory.
From each of these ten starting configurations at 800~K we restart independent 550~K
runs (Bussi thermostat~\cite{bussi_2007} with 0.1~ps time constant)
and monitor the formation of clusters
with different levels of \gls{vdW} corrections over 250~ps of \gls{MD}.
The results for \gls{TS} and \gls{lMBD} are shown in \fig{03}~(a). We see that, as the system
equilibrates, the \gls{lMBD} calculations decrease in potential energy by much more than the
\gls{TS} calculations. This is indicative of clustering, which we observe in all independent
\gls{lMBD} runs. By contrast, the \gls{TS} calculations retain the gaseous phase throughout
the dynamics.

Having established a significant qualitative difference in behavior between \gls{lMBD} and \gls{TS}
for white phosphorus at low density, we examine the ability of the \gls{cTS} approach to
reproduce the \gls{lMBD} results. Panel (b) of \fig{03} shows the potential energy results
for ten independent runs, applying parametrization updates every 10 and 100 steps. Both
calculations converge to similar, although slightly lower than the \gls{lMBD} reference,
potential energy ranges. The main
difference between these \gls{cTS} calculations and the reference \gls{lMBD} calculations is the
rate at which clustering takes place (faster in \gls{cTS}). We can see how the pathological
behavior observed in the \gls{TS} runs has been removed and the correct structure is reproduced.
\fig{03}~(c) further studies the structural details of the P$_4$ clusters through a \gls{RDF}
analysis. We observe slightly more structured clusters with \gls{cTS} (higher \gls{RDF}
values after the first minimum) but in much better agreement with the \gls{lMBD} result than
the \gls{TS} calculation, where the \gls{RDF} quickly plateaus to unity, the ideal value for
a homogeneous amorphous fluid. As we mentioned in Sec.~\ref{11}, we could manually tune the
\gls{TS} damping function parameters $d$ and $s_\text{R}$ to further improve
the agreement between \gls{cTS} and \gls{lMBD} for this example; however, these parameters are
not universally transferable and thus we opt here to retain the standard \gls{PBE}
parametrization~\cite{tkatchenko_2009} for the sake of better transferability.

\subsection{Scaling factor evolution and stability}

\begin{figure}
\centering
\includegraphics[width=\linewidth]{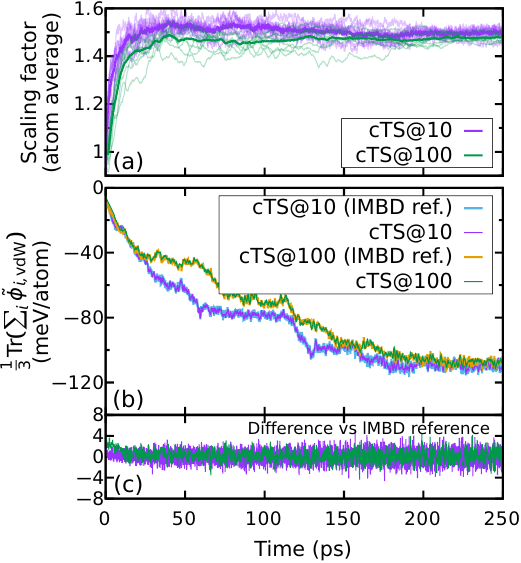}
\caption{(a) Time evolution of the cTS scaling factors for two different update frequencies in
the same systems of \fig{03}. (b) Evolution of the average trace of the damped virials during
the cTS runs, showing the lMBD targets. (c) Differences between instantaneous cTS and target
lMBD damped virials computed from (b), showing the initial equilibration time needed is longer for the
cTS run with less frequent reparametrization updates.}
\label{10}
\end{figure}

Finally, in \fig{10} we show the evolution of the different properties involved in the on-the-fly
\gls{TS} reparametrization for the same spontaneous clustering example of the previous section.
We choose $r_\text{min} = 5$~\AA{} and $r_\text{max} = 6$~\AA{} for the local virial damping
in \eq{08}, although we observed similar behavior also decreasing $r_\text{min}$ to 4~\AA{}.
In panel (a), we monitor the convergence of the
scaling factors for several independent runs (thin lines), as well as their average
(thick lines), for $N_\text{c} = 10$
and $N_\text{c} = 100$, equivalent to applying corrections every 20~fs and 200~fs,
respectively. We observe a faster convergence of the average scaling factor across
runs for $N_\text{c} = 10$, as well as significantly smaller scatter of the results
for individual runs. For $N_\text{c} = 100$, on the other hand, individual runs are
instantaneously less noisy (the local amplitude of the oscillations is smaller) but
they take longer to converge to the stationary solution and the variation across individual
runs is also larger. This is both due to the direct effect of less frequent updating
of the scaling factors and, indirectly, to the slower rate at which the stationary
structures (i.e., the clusters) are formed---recall that the effective $\tilde{C}_6$
parameters are environment-dependent and thus their value will continue to evolve
for as long as the atomistic structure continues to evolve.

In \fig{10}~(b) we show how the sum of the damped local \gls{vdW} virials in the
\gls{cTS} approximation closely follows the \gls{lMBD} reference, for clarity presented
for a single run (as averages across runs are meaningless in this case). With both tested
updating frequencies, $N_\text{c} = 10$ and $N_\text{c} = 100$, the instantaneous (damped)
virial closely matches the target value. Panel (c) shows the difference between
\gls{cTS} and \gls{lMBD} curves, highlighting how the instantaneous behavior for both
updating frequencies tested is very similar, except for an initial equilibration period
where it takes the $N_\text{c} = 100$ calculation of the order of 10~ps to catch on. This
is easily explained by the fact that at the beginning of the simulation, i.e., when the
correction scheme is switched on, the mismatch between the instantaneous scaling
factors ($s_i (t_{-1}) = 1$ for all $i$) and target scaling factors is largest. It takes
the calculation with slower correction the mentioned 10~ps to bridge this systematic
difference, after which the errors oscillate around zero. We note here in passing that
our implementation allows to ``dump'' to file and subsequently read the scaling
factors at the end of an \gls{MD} run so that the calculation can be restarted
with consistent atomic positions \textit{and} scaling factors.

\section{Phase diagram of white phosphorus}

\begin{figure}[t]
\begin{tabular}{c c}
\includegraphics[width=0.25\linewidth,trim={16cm 2cm 16cm 0}]{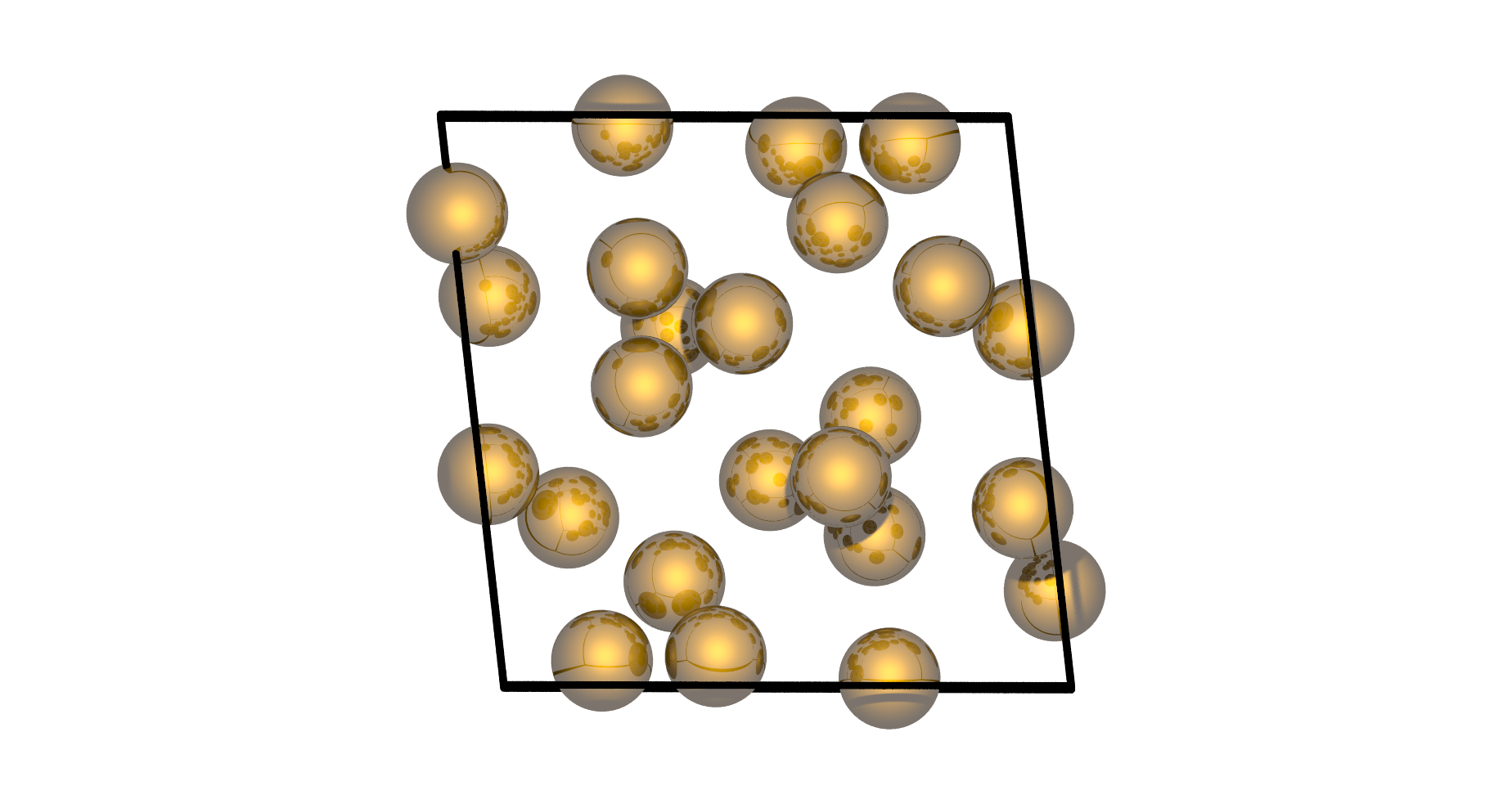}
&
\includegraphics[width=0.75\linewidth,trim={10cm 3cm 10cm 0}]{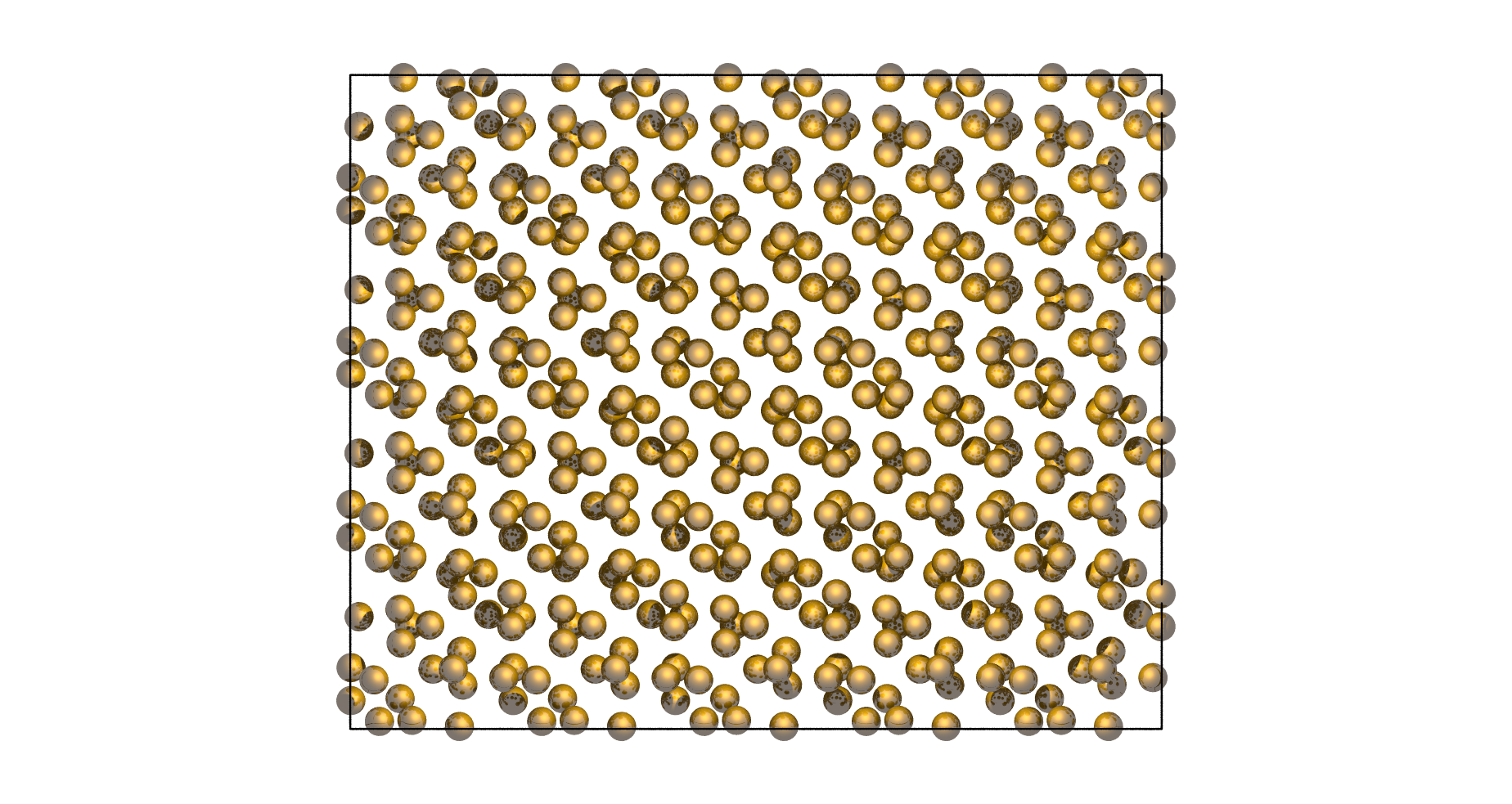}
\end{tabular}
\caption{(Left) Triclinic primitive unit cell of $\beta$-P$_4$. (Right)
Quasi-commensurate orthorhombic supercell of $\beta$-P$_4$ consisting of 120
primitive unit cells under slight strain.}
\label{quasi-commensurate}
\label{triclinic}
\end{figure}

As a benchmark of the proposed \gls{cTS} model, i.e., to validate the on-the-fly
reparametrization of dispersion interactions during \gls{MD}, we have calculated a
selected part of the metastable phase diagram of white phosphorus given in
Ref.~\onlinecite{clark2010compressibility}. Our objective is to reproduce 1) the
order-disorder transition and 2) the solid to molecular liquid transition. We do this
by running \gls{MD} simulations and monitoring the evolution with $P$ and $T$ of dynamical
properties (rotational and translational diffusivity parameters) and by comparing
the Gibbs free energy of the two phases. Our initial configuration corresponds to
the $\beta$-P$_{4}$ phase, which has a triclinic unit cell consisting of six
P$_4$ molecules with a space group of P$\bar{1}$~\cite{simon1987crystal} in
Hermann-Mauguin notation. The primitive unit cell is visualized in Fig.~\ref{triclinic} (left).
Because our \gls{DoS} analysis code is currently limited to orthorhombic unit cells,
and \gls{MD} simulation for large systems is generally more efficient in rectangular simulation
boxes, we created a quasi-commensurate supercell with 720 P$_4$ molecules that approximates the
crystal structure of the equivalent triclinic one. This supercell is visualized in
Fig.~\ref{quasi-commensurate} (right).

We ran \gls{MD} simulations with TurboGAP using cTS@100, meaning that the \gls{TS} parameters
were corrected by an \gls{lMBD} calculation every 100 time steps, and used a Verlet integration
time step of 2~fs for all calculations. While thermostatting is used for all simulations, we use
barostatting solely for the purpose of efficiently accessing different
regions of the phase diagram. Thus, the choice of barostat is not critical here
as all our analyses are carried out within the $NVT$ ensemble, and we employ the Berendsen
\textit{barostat}~\cite{berendsen_1984} for box rescaling with the three Cartesian dimensions
of the orthorhombic simulation box allowed to evolve independently. The choice of thermostat
should be handled more carefully precisely because we use the $NVT$ runs for analysis,
and so we avoid known issues with the Berendsen \textit{thermostat} by using the Bussi
stochastic velocity-rescaling approach~\cite{bussi_2007},
which is similarly efficient for equilibration but samples the correct (canonical) ensemble.

Our \gls{MD} protocol is as follows. The structure of the
quasi-commensurate cell was first equilibrated at $T = 1$~K with strong coupling to the
thermostat by setting the time coupling constant to 1~fs without a barostat until the
potential energy of the system had converged; this method of geometry optimization is often
referred to as damped \gls{MD}, and performs similarly to, e.g., gradient-descent geometry
optimization. After this initial equilibration at $T \approx 0$~K, we switched the time
coupling constant of the thermostat to 100 fs and switched on the barostat to generate
higher pressure structures at 1 K, starting from the last snapshot of the equilibration.
The barostat time coupling constant was set to 1000 fs and the bulk modulus parameter $\gamma_{\text{P}} = 2$ in units of the inverse compressibility of water:
$1/(4.5\times 10^{-5} \text{ bar}^{-1})$. This parameter is somewhat pressure-dependent
and was chosen by calculating the average for the solid: the parameter should be chosen such
that it is roughly equal to $\sqrt{P/1000}$, where $P$ is the pressure in units of bar.
We generated structures in the pressure range $[0, 12000]$ bar in 1000 bar intervals always
starting from the last snapshot of the previous pressure ramp-up. The interval is the region
of interest in Ref.~\onlinecite{clark2010compressibility} for the phase transitions we want
to observe. The increase in pressure was done in 20,000 \gls{MD} steps for each value.
This procedure at low temperature allows us to generate initial configurations at various
pressures, that retain their solid structure, for subsequent scans along the temperature
axis.

Thus, after generating the higher pressure structures, we started increasing the temperature for
each pressure snapshot. We first increased it to 100 K and then in 100 K increments to 1000 K for the 0 bar snapshot. Then, we increased the temperature to 150 K and then in 100 K increments to 950 K for the 1000 bar snapshot. The 2000 bar snapshot was again sampled on the $[100, 1000]$ K interval and so on, alternating up to the 12000 bar snapshot. This staggered temperature increase was used to better sample the $(P,T)$ space with less points. This way, we comprehensively
sample the temperature range $[0, 1000]$ K and generate the molecular liquid at high temperatures. We note that
the temperature at which spontaneous melting is observed will always overestimate the thermodynamic
phase transition temperature because of the existence of kinetic effects, i.e., free-energy
barriers that will not be overcome during the simulation times available to \gls{MD} unless the
simulations are carried out at higher temperatures.
The temperature increments were also done over 20,000 \gls{MD} steps. At even higher
temperatures, the structure starts to polymerize, creating polymeric liquid. We avoid this
temperature region as we need to keep the P$_4$ molecules intact for the molecular liquid phase
transition analysis. After the temperature ramp-up, we have our configurations for the solid
calculation (which include the spontaneously generated liquid at high temperature). We refer to
these scans as the ``solid series''. To generate
the configurations for the liquid series, we started decreasing the temperature from
1000~K (or 950~K depending on the pressure) in 100 K decrements, i.e., we scan the temperature range in the opposite direction.
Because of the rotational and translational disorder, the molecular liquid does not transition
back to crystalline $\beta$-P$_4$. We can analyze the solid and liquid series to determine
the region of phase transition based on the Gibbs free energies of both series, thus working around
the issue of metastability of the solid and liquid phases above and below the thermodynamic
melting temperature, respectively.

We then turned the barostat off and ran $NVT$ simulations for the last snapshots of the
configurations of the solid series and the liquid series.
We ran $NVT$ for 50,000 \gls{MD} steps with the same thermostat
parameters as in the structure generation. The average temperature, total energy, volume, and
pressure for the Gibbs free energy are obtained from the output of this calculation, and the
trajectory is analyzed with the DoSPT code~\cite{caro_2016,caro_2017b,ref_dospt}
to obtain the entropy and the translational
and rotational density of states with the 2PT formalism~\cite{lin_2003}. Within 2PT, the
kinetic energy of the system is decomposed into its translational, rotational and vibrational
\gls{DoS}. The zero-frequency \gls{DoS} gives the diffusivity of the system: the translational
zero-frequency \gls{DoS} is zero for the solid and finite for the liquid and related
to the self-diffusion coefficient of P$_4$ molecules; the rotational zero-frequency \gls{DoS}
is zero for the rotationally ordered solid (where the P$_4$ tetrahedra ``rock'' about their
equilibrium positions and orientations) and finite for the rotationally disordered solid and
the liquid (where the orientation of the P$_4$ tetrahedra is not fixed).

\begin{figure}[t]
  \includegraphics[width=\linewidth]{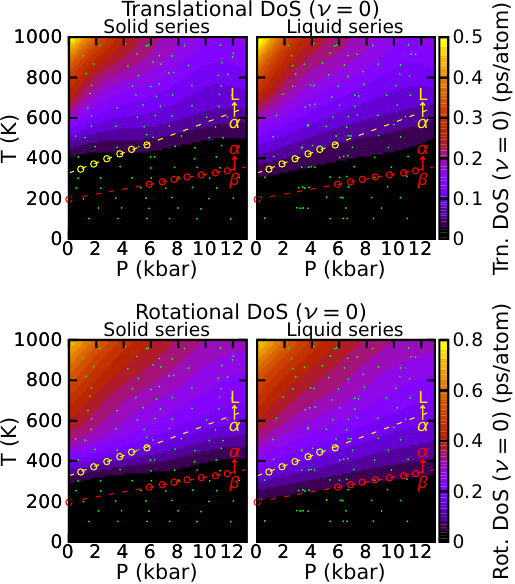}
  \caption{Zero-frequency density of states for the solid and liquid series of the
  configurations. The red lines indicate the (extrapolated from experiments in
  Ref.~\onlinecite{bridgman_1914}) position of the rotational order-disorder phase transition
  and the yellow line (also extrapolated from experimental values in
  Ref.~\onlinecite{bridgman_1914}) shows the phase boundary for the transition between solid
  white phosphorus and the molecular liquid. Warm colors indicate higher values and
  dark colors indicate lower values. The small green dots indicate the positions
  in ($P,T$) space of the collected simulation data.}
  \label{dos}
\end{figure}

\begin{figure}[t]
  \includegraphics[width=\linewidth]{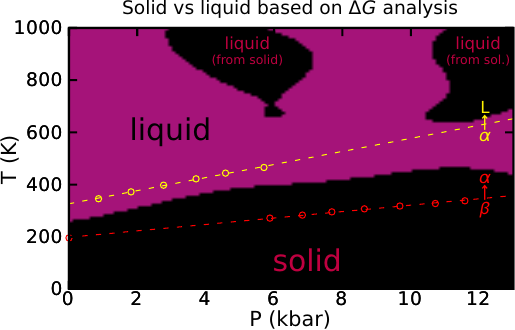}
  \caption{Metastable phase diagram of white phosphorus P$_4$ calculated using the
  solid and liquid series to determine which one has a lower Gibbs free energy at
  each point on the interpolated surface. The lines are the same as in Fig.~\ref{dos}
  and purple areas indicate that the liquid series has a lower Gibbs free energy at
  those points while the black areas indicate that the solid series configurations
  are lower in energy. The black areas at high temperature correspond to liquid
  configurations which were obtained by spontaneous melting from the solid.}
  \label{phase_diagram}
\end{figure}

After calculating the Gibbs free energies and the translational and rotational
zero-frequency \gls{DoS} at our pressure and temperature values, we used \gls{GPR}
to fit models of the data as a function of the two independent variables $P$ and $T$.
We use a Gaussian kernel, given by:
\begin{equation}
    k(P_i,P_j,T_i,T_j) = \exp\left( -\frac{1}{2}\left(\left(\frac{P_i-P_j}{\sigma_P}\right)^2 + \left(\frac{T_i-T_j}{\sigma_T}\right)^2 \right) \right).
\end{equation}
During training, each pressure-temperature pair $(P_i,T_i)$ is used to construct the
(dimensionless) covariance matrix $\mathbf{C}$ with elements:
\begin{equation}
    C(i,j) = k(P_i,P_j,T_i,T_j) + \lambda^2 \delta_{ij}.
\end{equation}
The model is regularized by adding a noise-to-signal parameter $\lambda^2$ to the diagonal
of the matrix to prevent overfitting. The hyperparameters of the model were optimized
using $n$-fold cross-validation. The optimal Gaussian smearing for the pressure was found to be
$\sigma_P = 5500$ bar and $\sigma_T = 220$ K for the temperature. The optimal regularization
parameter was found to be $\lambda = 0.05$, leading to errors below 1~meV/atom for the
Gibbs free energy. We can then solve for the fitting coefficients:
\begin{equation}
    \boldsymbol{\beta} = \mathbf{C}^{-1}\mathbf{y},
\end{equation}
where $\mathbf{y}$ contains either the calculated Gibbs free energies or the zero-frequency
\gls{DoS} for the solid or the liquid series. For the Gibbs free energy, a second-order
polynomial in $P$ and $T$ is first fit on the data via least-squares minimization and
subtracted before the \gls{GPR} model is trained, then added back during prediction.
This allows us to interpolate values for the surfaces at a new point $(P_*,T_*)$ by:
\begin{equation}
    y_*(P_*,T_*) = \sum\limits_{s=1}^M \beta_s k(P_*,P_s,T_*,T_s),
\end{equation}
where $M$ is the number of $(P,T)$-points we calculated using \gls{MD}, and $\beta_s$ are
elements of the coefficient vector $\boldsymbol{\beta}$. We proceed in this way because the
\gls{MD} data for the solid and liquid series are obtained for different ($P,T$) values and
cannot be compared directly. Thus, for the Gibbs free energy analysis,
we proceed by comparing the \gls{GPR} models trained for
the solid and liquid series instead of directly comparing the collected data.
In the case of the zero-frequency \gls{DoS}, the
\gls{GPR} models simply enable smoother interpolation between data points.

The interpolated translational and rotational zero-frequency \gls{DoS} surfaces for the
solid and liquid series are shown in Fig.~\ref{dos}. It can be seen that the onset of
the increase in the zero-frequency rotational \gls{DoS} in the solid series more or less
corresponds to the experimental $\beta-\alpha$ rotational order-disorder phase
transition~\cite{bridgman_1914,clark2010compressibility}. The increase in the zero-frequency
translational \gls{DoS} indicates that the crystal is starting to lose its structure and
turning into a liquid. This more or less agrees with the experimental phase transition
between crystalline white phosphorus and molecular P$_4$
liquid~\cite{bridgman_1914,clark2010compressibility}. For the liquid series, the
zero-frequency \gls{DoS} is higher around these values, indicating that there is
hysteresis in the phase transition, as expected and discussed earlier in the context of
free-energy barriers and phase metastability within \gls{MD} simulation time scales.

The phase diagram with the $\beta$-P$_4$ to molecular liquid phase boundary based
on the comparison of the interpolated Gibbs free energy surfaces is shown in Fig.~\ref{phase_diagram}. The purple regions indicate a lower Gibbs free energy
for the liquid series whereas the black area is where the solid series has a lower Gibbs
free energy. Both the solid series and the liquid series lead to liquid configurations
at high-enough temperatures so the apparent ``artifacts'' in the high-temperature region
of the plot are expected. Our simulations predict that the molecular liquid phase transition
happens at approximately 50~K lower temperature than the experimental
one~\cite{bridgman_1914,clark2010compressibility}, which is a reasonable agreement given
the small energy differences involved. In addition, the pressure dependence of the melting
temperature is correctly reproduced. These results could be refined by
calculating more $(P,T)$-points with \gls{MD}. The accuracy of the protocol used for the
free-energy analysis could also be assessed by comparing with other methods like coexistence
simulations~\cite{morris_2002} or nested sampling~\cite{partay_2021}. Our prior experience
indicates that small discrepancies can be expected between different methods~\cite{jana_2023}
and in some cases the reference method (i.e., PBE-DFT here) can be the limiting
factor in achieving quantitative agreement with experiment~\cite{kloppenburg_2023}. Furthermore, the quasi-commensurate structure used in the simulations may be slightly less stable than the triclinic structure of real $\beta$-P$_4$, causing it to melt at a lower temperature.

\section{Summary and conclusions}

In the present work we have developed a method for accelerated \gls{MD} simulations
with \gls{lMBD} corrections, which are necessary to achieve satisfactory accuracy
in some molecular and material systems. Despite being linear-scaling with the number of atoms,
\gls{lMBD} still suffers from a
significant and inevitable computational overhead compared to pairwise dispersion-correction
models. This means that, in calculations run with \gls{GAP} and other emerging \glspl{MLP},
high-level \gls{vdW} corrections would become a significant computational bottleneck when
needed. We have shown that our \gls{cTS} approach, an on-the-fly reparametrization of the
\gls{TS} model with \gls{lMBD}-accurate corrections, offers the desirable tradeoff between
accuracy and computational efficiency. By only computing the high-level (expensive)
corrections every $N_{\text{c}}$ steps and using the derived properties to parametrize
a computationally affordable force field, \gls{cTS} speeds up the calculation by almost a
factor of $N_{\text{c}}$ while retaining the needed accuracy for \gls{MD}.

We have benchmarked our method by calculating a part of the metastable phase diagram
of white phosphorus, where we expect accurate dispersion corrections to be essential for
meaningful results: the pairwise dispersion corrections are known to handle rotational
disorder poorly even in simple systems with rotations such as a benzene dimer~\cite{blood2016analytical,hermann2017first} so care should be taken when applying dispersion to such systems. In our results, we managed to demarcate the phase boundaries between the
rotationally ordered and disordered phases and the transition from crystalline molecular
solid to molecular liquid, getting remarkably close to the available experimental
data~\cite{bridgman_1914,clark2010compressibility}.
While these are promising proof of principle and application results for the proposed
on-the-fly reparametrization, we plan to implement this approach more rigorously in the
future. One idea is to optimize all of the parameters of the
\gls{TS} model fully consistently on the fly, instead of adding a correction to the energy
and pressure, and the proposed heuristic recipe for the forces based on the local virials.

\begin{acknowledgments}
The authors acknowledge financial support from the Research Council of Finland/Academy of Finland through grants numbers 321713 (H.M. and M.A.C.), 347252 (H.M. and M.A.C.) and 330488 (M.A.C.).
Innovation Study XCALE has received funding
through the Inno4scale project, which is funded by the European High-Performance Computing
Joint Undertaking (JU) under Grant Agreement No 101118139. The JU receives support from the
European Union's Horizon Europe Programme.
T.A-N. has been supported in part under Academy of Finland's grants to QTF Center of Excellence no. 31229 and European Union – NextGenerationEU instrument no. 353298.
Computational resources from CSC -- the Finnish IT Center for Science and Aalto University's Science-IT Project are gratefully acknowledged.
\end{acknowledgments}

\def\bibsection{}
\section*{References}


\begin{thebibliography}{83}%
\makeatletter
\providecommand \@ifxundefined [1]{%
 \@ifx{#1\undefined}
}%
\providecommand \@ifnum [1]{%
 \ifnum #1\expandafter \@firstoftwo
 \else \expandafter \@secondoftwo
 \fi
}%
\providecommand \@ifx [1]{%
 \ifx #1\expandafter \@firstoftwo
 \else \expandafter \@secondoftwo
 \fi
}%
\providecommand \natexlab [1]{#1}%
\providecommand \enquote  [1]{``#1''}%
\providecommand \bibnamefont  [1]{#1}%
\providecommand \bibfnamefont [1]{#1}%
\providecommand \citenamefont [1]{#1}%
\providecommand \href@noop [0]{\@secondoftwo}%
\providecommand \href [0]{\begingroup \@sanitize@url \@href}%
\providecommand \@href[1]{\@@startlink{#1}\@@href}%
\providecommand \@@href[1]{\endgroup#1\@@endlink}%
\providecommand \@sanitize@url [0]{\catcode `\\12\catcode `\$12\catcode
  `\&12\catcode `\#12\catcode `\^12\catcode `\_12\catcode `\%12\relax}%
\providecommand \@@startlink[1]{}%
\providecommand \@@endlink[0]{}%
\providecommand \url  [0]{\begingroup\@sanitize@url \@url }%
\providecommand \@url [1]{\endgroup\@href {#1}{\urlprefix }}%
\providecommand \urlprefix  [0]{URL }%
\providecommand \Eprint [0]{\href }%
\providecommand \doibase [0]{http://dx.doi.org/}%
\providecommand \selectlanguage [0]{\@gobble}%
\providecommand \bibinfo  [0]{\@secondoftwo}%
\providecommand \bibfield  [0]{\@secondoftwo}%
\providecommand \translation [1]{[#1]}%
\providecommand \BibitemOpen [0]{}%
\providecommand \bibitemStop [0]{}%
\providecommand \bibitemNoStop [0]{.\EOS\space}%
\providecommand \EOS [0]{\spacefactor3000\relax}%
\providecommand \BibitemShut  [1]{\csname bibitem#1\endcsname}%
\let\auto@bib@innerbib\@empty
\bibitem [{\citenamefont {Truhlar}(2019)}]{truhlar2019dispersion}%
  \BibitemOpen
  \bibfield  {author} {\bibinfo {author} {\bibfnamefont {D.~G.}\ \bibnamefont
  {Truhlar}},\ }\bibfield  {title} {\enquote {\bibinfo {title} {{Dispersion
  forces: Neither fluctuating nor dispersing}},}\ }\href@noop {} {\bibfield
  {journal} {\bibinfo  {journal} {J. Chem. Educ.}\ }\textbf {\bibinfo {volume}
  {96}},\ \bibinfo {pages} {1671} (\bibinfo {year} {2019})}\BibitemShut
  {NoStop}%
\bibitem [{\citenamefont {Blood-Forsythe}\ \emph {et~al.}(2016)\citenamefont
  {Blood-Forsythe}, \citenamefont {Markovich}, \citenamefont {DiStasio},
  \citenamefont {Car},\ and\ \citenamefont
  {Aspuru-Guzik}}]{blood2016analytical}%
  \BibitemOpen
  \bibfield  {author} {\bibinfo {author} {\bibfnamefont {M.~A.}\ \bibnamefont
  {Blood-Forsythe}}, \bibinfo {author} {\bibfnamefont {T.}~\bibnamefont
  {Markovich}}, \bibinfo {author} {\bibfnamefont {R.~A.}\ \bibnamefont
  {DiStasio}}, \bibinfo {author} {\bibfnamefont {R.}~\bibnamefont {Car}}, \
  and\ \bibinfo {author} {\bibfnamefont {A.}~\bibnamefont {Aspuru-Guzik}},\
  }\bibfield  {title} {\enquote {\bibinfo {title} {Analytical nuclear gradients
  for the range-separated many-body dispersion model of noncovalent
  interactions},}\ }\href@noop {} {\bibfield  {journal} {\bibinfo  {journal}
  {Chem. Sci.}\ }\textbf {\bibinfo {volume} {7}},\ \bibinfo {pages} {1712}
  (\bibinfo {year} {2016})}\BibitemShut {NoStop}%
\bibitem [{\citenamefont {Hermann}\ \emph {et~al.}(2017)\citenamefont
  {Hermann}, \citenamefont {DiStasio~Jr},\ and\ \citenamefont
  {Tkatchenko}}]{hermann2017first}%
  \BibitemOpen
  \bibfield  {author} {\bibinfo {author} {\bibfnamefont {J.}~\bibnamefont
  {Hermann}}, \bibinfo {author} {\bibfnamefont {R.~A.}\ \bibnamefont
  {DiStasio~Jr}}, \ and\ \bibinfo {author} {\bibfnamefont {A.}~\bibnamefont
  {Tkatchenko}},\ }\bibfield  {title} {\enquote {\bibinfo {title}
  {First-principles models for van der waals interactions in molecules and
  materials: Concepts, theory, and applications},}\ }\href@noop {} {\bibfield
  {journal} {\bibinfo  {journal} {Chem. Rev.}\ }\textbf {\bibinfo {volume}
  {117}},\ \bibinfo {pages} {4714} (\bibinfo {year} {2017})}\BibitemShut
  {NoStop}%
\bibitem [{\citenamefont {Becke}\ and\ \citenamefont
  {Johnson}(2005)}]{becke_2005}%
  \BibitemOpen
  \bibfield  {author} {\bibinfo {author} {\bibfnamefont {A.~D.}\ \bibnamefont
  {Becke}}\ and\ \bibinfo {author} {\bibfnamefont {E.~R.}\ \bibnamefont
  {Johnson}},\ }\bibfield  {title} {\enquote {\bibinfo {title} {Exchange-hole
  dipole moment and the dispersion interaction},}\ }\href@noop {} {\bibfield
  {journal} {\bibinfo  {journal} {J. Chem. Phys.}\ }\textbf {\bibinfo {volume}
  {122}},\ \bibinfo {pages} {154104} (\bibinfo {year} {2005})}\BibitemShut
  {NoStop}%
\bibitem [{\citenamefont {Tkatchenko}\ \emph {et~al.}(2012)\citenamefont
  {Tkatchenko}, \citenamefont {DiStasio~Jr}, \citenamefont {Car},\ and\
  \citenamefont {Scheffler}}]{tkatchenko_2012}%
  \BibitemOpen
  \bibfield  {author} {\bibinfo {author} {\bibfnamefont {A.}~\bibnamefont
  {Tkatchenko}}, \bibinfo {author} {\bibfnamefont {R.~A.}\ \bibnamefont
  {DiStasio~Jr}}, \bibinfo {author} {\bibfnamefont {R.}~\bibnamefont {Car}}, \
  and\ \bibinfo {author} {\bibfnamefont {M.}~\bibnamefont {Scheffler}},\
  }\bibfield  {title} {\enquote {\bibinfo {title} {Accurate and efficient
  method for many-body van der {Waals} interactions},}\ }\href@noop {}
  {\bibfield  {journal} {\bibinfo  {journal} {Phys. Rev. Lett.}\ }\textbf
  {\bibinfo {volume} {108}},\ \bibinfo {pages} {236402} (\bibinfo {year}
  {2012})}\BibitemShut {NoStop}%
\bibitem [{\citenamefont {Clark}\ and\ \citenamefont
  {Zaug}(2010)}]{clark2010compressibility}%
  \BibitemOpen
  \bibfield  {author} {\bibinfo {author} {\bibfnamefont {S.}~\bibnamefont
  {Clark}}\ and\ \bibinfo {author} {\bibfnamefont {J.}~\bibnamefont {Zaug}},\
  }\bibfield  {title} {\enquote {\bibinfo {title} {Compressibility of cubic
  white, orthorhombic black, rhombohedral black, and simple cubic black
  phosphorus},}\ }\href@noop {} {\bibfield  {journal} {\bibinfo  {journal}
  {Phys. Rev. B}\ }\textbf {\bibinfo {volume} {82}},\ \bibinfo {pages} {134111}
  (\bibinfo {year} {2010})}\BibitemShut {NoStop}%
\bibitem [{\citenamefont {Deringer}\ \emph
  {et~al.}(2020{\natexlab{a}})\citenamefont {Deringer}, \citenamefont
  {Pickard},\ and\ \citenamefont {Proserpio}}]{deringer_2020b}%
  \BibitemOpen
  \bibfield  {author} {\bibinfo {author} {\bibfnamefont {V.~L.}\ \bibnamefont
  {Deringer}}, \bibinfo {author} {\bibfnamefont {C.~J.}\ \bibnamefont
  {Pickard}}, \ and\ \bibinfo {author} {\bibfnamefont {D.~M.}\ \bibnamefont
  {Proserpio}},\ }\bibfield  {title} {\enquote {\bibinfo {title}
  {Hierarchically structured allotropes of phosphorus from data-driven
  exploration},}\ }\href@noop {} {\bibfield  {journal} {\bibinfo  {journal}
  {Angew. Chem. Int. Ed.}\ }\textbf {\bibinfo {volume} {59}},\ \bibinfo {pages}
  {15880} (\bibinfo {year} {2020}{\natexlab{a}})}\BibitemShut {NoStop}%
\bibitem [{\citenamefont {Brown}\ and\ \citenamefont
  {Rundqvist}(1965)}]{brown1965refinement}%
  \BibitemOpen
  \bibfield  {author} {\bibinfo {author} {\bibfnamefont {A.}~\bibnamefont
  {Brown}}\ and\ \bibinfo {author} {\bibfnamefont {S.}~\bibnamefont
  {Rundqvist}},\ }\bibfield  {title} {\enquote {\bibinfo {title} {Refinement of
  the crystal structure of black phosphorus},}\ }\href@noop {} {\bibfield
  {journal} {\bibinfo  {journal} {Acta Crystallogr.}\ }\textbf {\bibinfo
  {volume} {19}},\ \bibinfo {pages} {684} (\bibinfo {year} {1965})}\BibitemShut
  {NoStop}%
\bibitem [{\citenamefont {Jamieson}(1963)}]{jamieson1963crystal}%
  \BibitemOpen
  \bibfield  {author} {\bibinfo {author} {\bibfnamefont {J.}~\bibnamefont
  {Jamieson}},\ }\bibfield  {title} {\enquote {\bibinfo {title} {Crystal
  structures adopted by black phosphorus at high pressures},}\ }\href@noop {}
  {\bibfield  {journal} {\bibinfo  {journal} {Science}\ }\textbf {\bibinfo
  {volume} {139}},\ \bibinfo {pages} {1291} (\bibinfo {year}
  {1963})}\BibitemShut {NoStop}%
\bibitem [{\citenamefont {Kikegawa}\ and\ \citenamefont
  {Iwasaki}(1983)}]{kikegawa1983x}%
  \BibitemOpen
  \bibfield  {author} {\bibinfo {author} {\bibfnamefont {T.}~\bibnamefont
  {Kikegawa}}\ and\ \bibinfo {author} {\bibfnamefont {H.}~\bibnamefont
  {Iwasaki}},\ }\bibfield  {title} {\enquote {\bibinfo {title} {{An X-ray
  diffraction study of lattice compression and phase transition of crystalline
  phosphorus}},}\ }\href@noop {} {\bibfield  {journal} {\bibinfo  {journal}
  {Acta Crystallogr. B}\ }\textbf {\bibinfo {volume} {39}},\ \bibinfo {pages}
  {158} (\bibinfo {year} {1983})}\BibitemShut {NoStop}%
\bibitem [{\citenamefont {Akahama}\ \emph {et~al.}(1999)\citenamefont
  {Akahama}, \citenamefont {Kobayashi},\ and\ \citenamefont
  {Kawamura}}]{akahama1999simple}%
  \BibitemOpen
  \bibfield  {author} {\bibinfo {author} {\bibfnamefont {Y.}~\bibnamefont
  {Akahama}}, \bibinfo {author} {\bibfnamefont {M.}~\bibnamefont {Kobayashi}},
  \ and\ \bibinfo {author} {\bibfnamefont {H.}~\bibnamefont {Kawamura}},\
  }\bibfield  {title} {\enquote {\bibinfo {title}
  {Simple-cubic--simple-hexagonal transition in phosphorus under pressure},}\
  }\href@noop {} {\bibfield  {journal} {\bibinfo  {journal} {Phys. Rev. B}\
  }\textbf {\bibinfo {volume} {59}},\ \bibinfo {pages} {8520} (\bibinfo {year}
  {1999})}\BibitemShut {NoStop}%
\bibitem [{\citenamefont {Akahama}\ \emph {et~al.}(2000)\citenamefont
  {Akahama}, \citenamefont {Kawamura}, \citenamefont {Carlson}, \citenamefont
  {Le~Bihan},\ and\ \citenamefont {H{\"a}usermann}}]{akahama2000structural}%
  \BibitemOpen
  \bibfield  {author} {\bibinfo {author} {\bibfnamefont {Y.}~\bibnamefont
  {Akahama}}, \bibinfo {author} {\bibfnamefont {H.}~\bibnamefont {Kawamura}},
  \bibinfo {author} {\bibfnamefont {S.}~\bibnamefont {Carlson}}, \bibinfo
  {author} {\bibfnamefont {T.}~\bibnamefont {Le~Bihan}}, \ and\ \bibinfo
  {author} {\bibfnamefont {D.}~\bibnamefont {H{\"a}usermann}},\ }\bibfield
  {title} {\enquote {\bibinfo {title} {{Structural stability and equation of
  state of simple-hexagonal phosphorus to 280~{GPa}: Phase transition at
  262~{GPa}}},}\ }\href@noop {} {\bibfield  {journal} {\bibinfo  {journal}
  {Phys. Rev. B}\ }\textbf {\bibinfo {volume} {61}},\ \bibinfo {pages} {3139}
  (\bibinfo {year} {2000})}\BibitemShut {NoStop}%
\bibitem [{\citenamefont {Ruck}\ \emph {et~al.}(2005)\citenamefont {Ruck},
  \citenamefont {Hoppe}, \citenamefont {Wahl}, \citenamefont {Simon},
  \citenamefont {Wang},\ and\ \citenamefont {Seifert}}]{ruck2005fibrous}%
  \BibitemOpen
  \bibfield  {author} {\bibinfo {author} {\bibfnamefont {M.}~\bibnamefont
  {Ruck}}, \bibinfo {author} {\bibfnamefont {D.}~\bibnamefont {Hoppe}},
  \bibinfo {author} {\bibfnamefont {B.}~\bibnamefont {Wahl}}, \bibinfo {author}
  {\bibfnamefont {P.}~\bibnamefont {Simon}}, \bibinfo {author} {\bibfnamefont
  {Y.}~\bibnamefont {Wang}}, \ and\ \bibinfo {author} {\bibfnamefont
  {G.}~\bibnamefont {Seifert}},\ }\bibfield  {title} {\enquote {\bibinfo
  {title} {Fibrous red phosphorus},}\ }\href@noop {} {\bibfield  {journal}
  {\bibinfo  {journal} {Angew. Chem. Int. Ed.}\ }\textbf {\bibinfo {volume}
  {44}},\ \bibinfo {pages} {7616} (\bibinfo {year} {2005})}\BibitemShut
  {NoStop}%
\bibitem [{\citenamefont {Zhang}\ \emph {et~al.}(2017)\citenamefont {Zhang},
  \citenamefont {Xing}, \citenamefont {Li},\ and\ \citenamefont
  {Yan}}]{zhang2017hittorf}%
  \BibitemOpen
  \bibfield  {author} {\bibinfo {author} {\bibfnamefont {Z.}~\bibnamefont
  {Zhang}}, \bibinfo {author} {\bibfnamefont {D.}~\bibnamefont {Xing}},
  \bibinfo {author} {\bibfnamefont {J.}~\bibnamefont {Li}}, \ and\ \bibinfo
  {author} {\bibfnamefont {Q.}~\bibnamefont {Yan}},\ }\bibfield  {title}
  {\enquote {\bibinfo {title} {Hittorf's phosphorus: the missing link during
  transformation of red phosphorus to black phosphorus},}\ }\href@noop {}
  {\bibfield  {journal} {\bibinfo  {journal} {CrystEngComm}\ }\textbf {\bibinfo
  {volume} {19}},\ \bibinfo {pages} {905} (\bibinfo {year} {2017})}\BibitemShut
  {NoStop}%
\bibitem [{\citenamefont {Zhou}\ \emph {et~al.}(2023)\citenamefont {Zhou},
  \citenamefont {Elliott},\ and\ \citenamefont {Deringer}}]{zhou_2023}%
  \BibitemOpen
  \bibfield  {author} {\bibinfo {author} {\bibfnamefont {Y.}~\bibnamefont
  {Zhou}}, \bibinfo {author} {\bibfnamefont {S.~R.}\ \bibnamefont {Elliott}}, \
  and\ \bibinfo {author} {\bibfnamefont {V.~L.}\ \bibnamefont {Deringer}},\
  }\bibfield  {title} {\enquote {\bibinfo {title} {Structure and bonding in
  amorphous red phosphorus},}\ }\href@noop {} {\bibfield  {journal} {\bibinfo
  {journal} {Angew. Chem. Int. Ed.}\ }\textbf {\bibinfo {volume} {62}},\
  \bibinfo {pages} {e202216658} (\bibinfo {year} {2023})}\BibitemShut {NoStop}%
\bibitem [{\citenamefont {Simon}\ \emph {et~al.}(1987)\citenamefont {Simon},
  \citenamefont {Borrmann},\ and\ \citenamefont {Craubner}}]{simon1987crystal}%
  \BibitemOpen
  \bibfield  {author} {\bibinfo {author} {\bibfnamefont {A.}~\bibnamefont
  {Simon}}, \bibinfo {author} {\bibfnamefont {H.}~\bibnamefont {Borrmann}}, \
  and\ \bibinfo {author} {\bibfnamefont {H.}~\bibnamefont {Craubner}},\
  }\bibfield  {title} {\enquote {\bibinfo {title} {Crystal structure of ordered
  white phosphorus {($\beta$-P)}},}\ }\href@noop {} {\bibfield  {journal}
  {\bibinfo  {journal} {Phosphorus Sulfur}\ }\textbf {\bibinfo {volume} {30}},\
  \bibinfo {pages} {507} (\bibinfo {year} {1987})}\BibitemShut {NoStop}%
\bibitem [{\citenamefont {Simon}\ \emph {et~al.}(1997)\citenamefont {Simon},
  \citenamefont {Borrmann},\ and\ \citenamefont
  {Horakh}}]{simon1997polymorphism}%
  \BibitemOpen
  \bibfield  {author} {\bibinfo {author} {\bibfnamefont {A.}~\bibnamefont
  {Simon}}, \bibinfo {author} {\bibfnamefont {H.}~\bibnamefont {Borrmann}}, \
  and\ \bibinfo {author} {\bibfnamefont {J.}~\bibnamefont {Horakh}},\
  }\bibfield  {title} {\enquote {\bibinfo {title} {On the polymorphism of white
  phosphorus},}\ }\href@noop {} {\bibfield  {journal} {\bibinfo  {journal}
  {Chem. Ber.}\ }\textbf {\bibinfo {volume} {130}},\ \bibinfo {pages} {1235}
  (\bibinfo {year} {1997})}\BibitemShut {NoStop}%
\bibitem [{\citenamefont {Okudera}\ \emph {et~al.}(2005)\citenamefont
  {Okudera}, \citenamefont {Dinnebier},\ and\ \citenamefont
  {Simon}}]{okudera2005crystal}%
  \BibitemOpen
  \bibfield  {author} {\bibinfo {author} {\bibfnamefont {H.}~\bibnamefont
  {Okudera}}, \bibinfo {author} {\bibfnamefont {R.}~\bibnamefont {Dinnebier}},
  \ and\ \bibinfo {author} {\bibfnamefont {A.}~\bibnamefont {Simon}},\
  }\bibfield  {title} {\enquote {\bibinfo {title} {The crystal structure of
  {$\gamma$-P$_4$}, a low temperature modification of white phosphorus},}\
  }\href@noop {} {\bibfield  {journal} {\bibinfo  {journal} {Z. Kristallogr.
  Cryst. Mater.}\ }\textbf {\bibinfo {volume} {220}},\ \bibinfo {pages} {259}
  (\bibinfo {year} {2005})}\BibitemShut {NoStop}%
\bibitem [{\citenamefont {Katayama}\ \emph {et~al.}(2000)\citenamefont
  {Katayama}, \citenamefont {Mizutani}, \citenamefont {Utsumi}, \citenamefont
  {Shimomura}, \citenamefont {Yamakata},\ and\ \citenamefont
  {Funakoshi}}]{katayama2000first}%
  \BibitemOpen
  \bibfield  {author} {\bibinfo {author} {\bibfnamefont {Y.}~\bibnamefont
  {Katayama}}, \bibinfo {author} {\bibfnamefont {T.}~\bibnamefont {Mizutani}},
  \bibinfo {author} {\bibfnamefont {W.}~\bibnamefont {Utsumi}}, \bibinfo
  {author} {\bibfnamefont {O.}~\bibnamefont {Shimomura}}, \bibinfo {author}
  {\bibfnamefont {M.}~\bibnamefont {Yamakata}}, \ and\ \bibinfo {author}
  {\bibfnamefont {K.}~\bibnamefont {Funakoshi}},\ }\bibfield  {title} {\enquote
  {\bibinfo {title} {A first-order liquid--liquid phase transition in
  phosphorus},}\ }\href@noop {} {\bibfield  {journal} {\bibinfo  {journal}
  {Nature}\ }\textbf {\bibinfo {volume} {403}},\ \bibinfo {pages} {170}
  (\bibinfo {year} {2000})}\BibitemShut {NoStop}%
\bibitem [{\citenamefont {Deringer}\ \emph
  {et~al.}(2020{\natexlab{b}})\citenamefont {Deringer}, \citenamefont {Caro},\
  and\ \citenamefont {Cs{\'a}nyi}}]{deringer_2020}%
  \BibitemOpen
  \bibfield  {author} {\bibinfo {author} {\bibfnamefont {V.~L.}\ \bibnamefont
  {Deringer}}, \bibinfo {author} {\bibfnamefont {M.~A.}\ \bibnamefont {Caro}},
  \ and\ \bibinfo {author} {\bibfnamefont {G.}~\bibnamefont {Cs{\'a}nyi}},\
  }\bibfield  {title} {\enquote {\bibinfo {title} {A general-purpose
  machine-learning force field for bulk and nanostructured phosphorus},}\
  }\href@noop {} {\bibfield  {journal} {\bibinfo  {journal} {Nat. Commun.}\
  }\textbf {\bibinfo {volume} {11}},\ \bibinfo {pages} {1} (\bibinfo {year}
  {2020}{\natexlab{b}})}\BibitemShut {NoStop}%
\bibitem [{\citenamefont {Muhli}\ \emph {et~al.}(2024)\citenamefont {Muhli},
  \citenamefont {Ala-Nissila},\ and\ \citenamefont {Caro}}]{muhli_2024}%
  \BibitemOpen
  \bibfield  {author} {\bibinfo {author} {\bibfnamefont {H.}~\bibnamefont
  {Muhli}}, \bibinfo {author} {\bibfnamefont {T.}~\bibnamefont {Ala-Nissila}},
  \ and\ \bibinfo {author} {\bibfnamefont {M.~A.}\ \bibnamefont {Caro}},\
  }\bibfield  {title} {\enquote {\bibinfo {title} {Atom-wise formulation of the
  many-body dispersion problem for linear-scaling van der {Waals}
  corrections},}\ }\href@noop {} {\bibfield  {journal} {\bibinfo  {journal}
  {arXiv:2407.06409}\ } (\bibinfo {year} {2024})}\BibitemShut {NoStop}%
\bibitem [{\citenamefont {Bart{\'o}k}\ \emph {et~al.}(2010)\citenamefont
  {Bart{\'o}k}, \citenamefont {Payne}, \citenamefont {Kondor},\ and\
  \citenamefont {Cs{\'a}nyi}}]{bartok2010gaussian}%
  \BibitemOpen
  \bibfield  {author} {\bibinfo {author} {\bibfnamefont {A.~P.}\ \bibnamefont
  {Bart{\'o}k}}, \bibinfo {author} {\bibfnamefont {M.~C.}\ \bibnamefont
  {Payne}}, \bibinfo {author} {\bibfnamefont {R.}~\bibnamefont {Kondor}}, \
  and\ \bibinfo {author} {\bibfnamefont {G.}~\bibnamefont {Cs{\'a}nyi}},\
  }\bibfield  {title} {\enquote {\bibinfo {title} {Gaussian approximation
  potentials: The accuracy of quantum mechanics, without the electrons},}\
  }\href@noop {} {\bibfield  {journal} {\bibinfo  {journal} {Phys. Rev. Lett.}\
  }\textbf {\bibinfo {volume} {104}},\ \bibinfo {pages} {136403} (\bibinfo
  {year} {2010})}\BibitemShut {NoStop}%
\bibitem [{\citenamefont {Klawohn}\ \emph {et~al.}(2023)\citenamefont
  {Klawohn}, \citenamefont {Darby}, \citenamefont {Kermode}, \citenamefont
  {Cs\'anyi}, \citenamefont {Caro},\ and\ \citenamefont
  {Bart\'ok}}]{klawohn_2023}%
  \BibitemOpen
  \bibfield  {author} {\bibinfo {author} {\bibfnamefont {S.}~\bibnamefont
  {Klawohn}}, \bibinfo {author} {\bibfnamefont {J.~P.}\ \bibnamefont {Darby}},
  \bibinfo {author} {\bibfnamefont {J.~R.}\ \bibnamefont {Kermode}}, \bibinfo
  {author} {\bibfnamefont {G.}~\bibnamefont {Cs\'anyi}}, \bibinfo {author}
  {\bibfnamefont {M.~A.}\ \bibnamefont {Caro}}, \ and\ \bibinfo {author}
  {\bibfnamefont {A.~P.}\ \bibnamefont {Bart\'ok}},\ }\bibfield  {title}
  {\enquote {\bibinfo {title} {Gaussian approximation potentials: theory,
  software implementation and application examples},}\ }\href@noop {}
  {\bibfield  {journal} {\bibinfo  {journal} {J. Chem. Phys.}\ }\textbf
  {\bibinfo {volume} {159}},\ \bibinfo {pages} {174108} (\bibinfo {year}
  {2023})}\BibitemShut {NoStop}%
\bibitem [{\citenamefont {Tkatchenko}\ and\ \citenamefont
  {Scheffler}(2009)}]{tkatchenko_2009}%
  \BibitemOpen
  \bibfield  {author} {\bibinfo {author} {\bibfnamefont {A.}~\bibnamefont
  {Tkatchenko}}\ and\ \bibinfo {author} {\bibfnamefont {M.}~\bibnamefont
  {Scheffler}},\ }\bibfield  {title} {\enquote {\bibinfo {title} {Accurate
  molecular van der {Waals} interactions from ground-state electron density and
  free-atom reference data},}\ }\href@noop {} {\bibfield  {journal} {\bibinfo
  {journal} {Phys. Rev. Lett.}\ }\textbf {\bibinfo {volume} {102}},\ \bibinfo
  {pages} {073005} (\bibinfo {year} {2009})}\BibitemShut {NoStop}%
\bibitem [{\citenamefont {Behler}\ and\ \citenamefont
  {Parrinello}(2007)}]{behler_2007}%
  \BibitemOpen
  \bibfield  {author} {\bibinfo {author} {\bibfnamefont {J.}~\bibnamefont
  {Behler}}\ and\ \bibinfo {author} {\bibfnamefont {M.}~\bibnamefont
  {Parrinello}},\ }\bibfield  {title} {\enquote {\bibinfo {title} {Generalized
  neural-network representation of high-dimensional potential-energy
  surfaces},}\ }\href@noop {} {\bibfield  {journal} {\bibinfo  {journal} {Phys.
  Rev. Lett.}\ }\textbf {\bibinfo {volume} {98}},\ \bibinfo {pages} {146401}
  (\bibinfo {year} {2007})}\BibitemShut {NoStop}%
\bibitem [{\citenamefont {Deringer}\ \emph {et~al.}(2019)\citenamefont
  {Deringer}, \citenamefont {Caro},\ and\ \citenamefont
  {Cs{\'a}nyi}}]{deringer_2019}%
  \BibitemOpen
  \bibfield  {author} {\bibinfo {author} {\bibfnamefont {V.~L.}\ \bibnamefont
  {Deringer}}, \bibinfo {author} {\bibfnamefont {M.~A.}\ \bibnamefont {Caro}},
  \ and\ \bibinfo {author} {\bibfnamefont {G.}~\bibnamefont {Cs{\'a}nyi}},\
  }\bibfield  {title} {\enquote {\bibinfo {title} {Machine learning interatomic
  potentials as emerging tools for materials science},}\ }\href@noop {}
  {\bibfield  {journal} {\bibinfo  {journal} {Adv. Mater.}\ }\textbf {\bibinfo
  {volume} {31}},\ \bibinfo {pages} {1902765} (\bibinfo {year}
  {2019})}\BibitemShut {NoStop}%
\bibitem [{\citenamefont {Hellstr\"om}\ and\ \citenamefont
  {Behler}(2020)}]{hellstrom_2020}%
  \BibitemOpen
  \bibfield  {author} {\bibinfo {author} {\bibfnamefont {M.}~\bibnamefont
  {Hellstr\"om}}\ and\ \bibinfo {author} {\bibfnamefont {J.}~\bibnamefont
  {Behler}},\ }\enquote {\bibinfo {title} {High-dimensional neural network
  potentials for atomistic simulations},}\ in\ \href@noop {} {\emph {\bibinfo
  {booktitle} {Machine learning meets quantum physics}}}\ (\bibinfo
  {publisher} {Springer},\ \bibinfo {address} {Cham, Switzerland},\ \bibinfo
  {year} {2020})\ p.\ \bibinfo {pages} {253}\BibitemShut {NoStop}%
\bibitem [{\citenamefont {Sch\"utt}\ \emph {et~al.}(2020)\citenamefont
  {Sch\"utt}, \citenamefont {Chmiela}, \citenamefont {von Lilienfeld},
  \citenamefont {Tkatchenko}, \citenamefont {Tsuda},\ and\ \citenamefont
  {M\"uller}}]{schutt_2020}%
  \BibitemOpen
  \bibfield  {author} {\bibinfo {author} {\bibfnamefont {K.~T.}\ \bibnamefont
  {Sch\"utt}}, \bibinfo {author} {\bibfnamefont {S.}~\bibnamefont {Chmiela}},
  \bibinfo {author} {\bibfnamefont {O.~A.}\ \bibnamefont {von Lilienfeld}},
  \bibinfo {author} {\bibfnamefont {A.}~\bibnamefont {Tkatchenko}}, \bibinfo
  {author} {\bibfnamefont {K.}~\bibnamefont {Tsuda}}, \ and\ \bibinfo {author}
  {\bibfnamefont {K.-R.}\ \bibnamefont {M\"uller}},\ }\href@noop {} {\emph
  {\bibinfo {title} {Machine learning meets quantum physics}}},\ \bibinfo
  {edition} {1st}\ ed.\ (\bibinfo  {publisher} {Springer},\ \bibinfo {address}
  {Cham, Switzerland},\ \bibinfo {year} {2020})\BibitemShut {NoStop}%
\bibitem [{\citenamefont {Deringer}\ \emph {et~al.}(2021)\citenamefont
  {Deringer}, \citenamefont {Bart{\'o}k}, \citenamefont {Bernstein},
  \citenamefont {Wilkins}, \citenamefont {Ceriotti},\ and\ \citenamefont
  {Cs{\'a}nyi}}]{deringer_2021}%
  \BibitemOpen
  \bibfield  {author} {\bibinfo {author} {\bibfnamefont {V.~L.}\ \bibnamefont
  {Deringer}}, \bibinfo {author} {\bibfnamefont {A.~P.}\ \bibnamefont
  {Bart{\'o}k}}, \bibinfo {author} {\bibfnamefont {N.}~\bibnamefont
  {Bernstein}}, \bibinfo {author} {\bibfnamefont {D.~M.}\ \bibnamefont
  {Wilkins}}, \bibinfo {author} {\bibfnamefont {M.}~\bibnamefont {Ceriotti}}, \
  and\ \bibinfo {author} {\bibfnamefont {G.}~\bibnamefont {Cs{\'a}nyi}},\
  }\bibfield  {title} {\enquote {\bibinfo {title} {Gaussian process regression
  for materials and molecules},}\ }\href@noop {} {\bibfield  {journal}
  {\bibinfo  {journal} {Chem. Rev.}\ }\textbf {\bibinfo {volume} {121}},\
  \bibinfo {pages} {10073} (\bibinfo {year} {2021})}\BibitemShut {NoStop}%
\bibitem [{\citenamefont {Bart{\'o}k}\ \emph {et~al.}(2013)\citenamefont
  {Bart{\'o}k}, \citenamefont {Kondor},\ and\ \citenamefont
  {Cs{\'a}nyi}}]{bartok2013representing}%
  \BibitemOpen
  \bibfield  {author} {\bibinfo {author} {\bibfnamefont {A.~P.}\ \bibnamefont
  {Bart{\'o}k}}, \bibinfo {author} {\bibfnamefont {R.}~\bibnamefont {Kondor}},
  \ and\ \bibinfo {author} {\bibfnamefont {G.}~\bibnamefont {Cs{\'a}nyi}},\
  }\bibfield  {title} {\enquote {\bibinfo {title} {On representing chemical
  environments},}\ }\href@noop {} {\bibfield  {journal} {\bibinfo  {journal}
  {Phys. Rev. B}\ }\textbf {\bibinfo {volume} {87}},\ \bibinfo {pages} {184115}
  (\bibinfo {year} {2013})}\BibitemShut {NoStop}%
\bibitem [{\citenamefont {Deringer}\ and\ \citenamefont
  {Cs{\'a}nyi}(2017)}]{deringer_2017}%
  \BibitemOpen
  \bibfield  {author} {\bibinfo {author} {\bibfnamefont {V.~L.}\ \bibnamefont
  {Deringer}}\ and\ \bibinfo {author} {\bibfnamefont {G.}~\bibnamefont
  {Cs{\'a}nyi}},\ }\bibfield  {title} {\enquote {\bibinfo {title} {Machine
  learning based interatomic potential for amorphous carbon},}\ }\href@noop {}
  {\bibfield  {journal} {\bibinfo  {journal} {Phys. Rev. B}\ }\textbf {\bibinfo
  {volume} {95}},\ \bibinfo {pages} {094203} (\bibinfo {year}
  {2017})}\BibitemShut {NoStop}%
\bibitem [{\citenamefont {Muhli}\ \emph {et~al.}(2021)\citenamefont {Muhli},
  \citenamefont {Chen}, \citenamefont {Bart{\'o}k}, \citenamefont
  {Hern{\'a}ndez-Le{\'o}n}, \citenamefont {Cs{\'a}nyi}, \citenamefont
  {Ala-Nissila},\ and\ \citenamefont {Caro}}]{muhli2021machine}%
  \BibitemOpen
  \bibfield  {author} {\bibinfo {author} {\bibfnamefont {H.}~\bibnamefont
  {Muhli}}, \bibinfo {author} {\bibfnamefont {X.}~\bibnamefont {Chen}},
  \bibinfo {author} {\bibfnamefont {A.~P.}\ \bibnamefont {Bart{\'o}k}},
  \bibinfo {author} {\bibfnamefont {P.}~\bibnamefont {Hern{\'a}ndez-Le{\'o}n}},
  \bibinfo {author} {\bibfnamefont {G.}~\bibnamefont {Cs{\'a}nyi}}, \bibinfo
  {author} {\bibfnamefont {T.}~\bibnamefont {Ala-Nissila}}, \ and\ \bibinfo
  {author} {\bibfnamefont {M.~A.}\ \bibnamefont {Caro}},\ }\bibfield  {title}
  {\enquote {\bibinfo {title} {{Machine learning force fields based on local
  parametrization of dispersion interactions: Application to the phase diagram
  of C$_{60}$}},}\ }\href@noop {} {\bibfield  {journal} {\bibinfo  {journal}
  {Phys. Rev. B}\ }\textbf {\bibinfo {volume} {104}},\ \bibinfo {pages}
  {054106} (\bibinfo {year} {2021})}\BibitemShut {NoStop}%
\bibitem [{\citenamefont {Caro}(2019)}]{caro_2019}%
  \BibitemOpen
  \bibfield  {author} {\bibinfo {author} {\bibfnamefont {M.~A.}\ \bibnamefont
  {Caro}},\ }\bibfield  {title} {\enquote {\bibinfo {title} {Optimizing
  many-body atomic descriptors for enhanced computational performance of
  machine learning based interatomic potentials},}\ }\href@noop {} {\bibfield
  {journal} {\bibinfo  {journal} {Phys. Rev. B}\ }\textbf {\bibinfo {volume}
  {100}},\ \bibinfo {pages} {024112} (\bibinfo {year} {2019})}\BibitemShut
  {NoStop}%
\bibitem [{\citenamefont {Caro}\ \emph {et~al.}()\citenamefont {Caro} \emph
  {et~al.}}]{ref_turbogap}%
  \BibitemOpen
  \bibfield  {author} {\bibinfo {author} {\bibfnamefont {M.~A.}\ \bibnamefont
  {Caro}} \emph {et~al.},\ }\href@noop {} {\enquote {\bibinfo {title}
  {{TurboGAP}: Data-driven atomistic simulations},}\ }\bibinfo {howpublished}
  {\url{http://turbogap.fi}},\ \bibinfo {note} {accessed:
  2024-10-14}\BibitemShut {NoStop}%
\bibitem [{\citenamefont {Perdew}\ \emph {et~al.}(1996)\citenamefont {Perdew},
  \citenamefont {Burke},\ and\ \citenamefont
  {Ernzerhof}}]{perdew1996generalized}%
  \BibitemOpen
  \bibfield  {author} {\bibinfo {author} {\bibfnamefont {J.}~\bibnamefont
  {Perdew}}, \bibinfo {author} {\bibfnamefont {K.}~\bibnamefont {Burke}}, \
  and\ \bibinfo {author} {\bibfnamefont {M.}~\bibnamefont {Ernzerhof}},\
  }\bibfield  {title} {\enquote {\bibinfo {title} {Generalized gradient
  approximation made simple},}\ }\href@noop {} {\bibfield  {journal} {\bibinfo
  {journal} {Phys. Rev. Lett.}\ }\textbf {\bibinfo {volume} {77}},\ \bibinfo
  {pages} {3865} (\bibinfo {year} {1996})}\BibitemShut {NoStop}%
\bibitem [{\citenamefont {Kresse}\ and\ \citenamefont
  {Furthm{\"u}ller}(1996{\natexlab{a}})}]{kresse1996efficient}%
  \BibitemOpen
  \bibfield  {author} {\bibinfo {author} {\bibfnamefont {G.}~\bibnamefont
  {Kresse}}\ and\ \bibinfo {author} {\bibfnamefont {J.}~\bibnamefont
  {Furthm{\"u}ller}},\ }\bibfield  {title} {\enquote {\bibinfo {title}
  {Efficient iterative schemes for \textit{ab initio} total-energy calculations
  using a plane-wave basis set},}\ }\href@noop {} {\bibfield  {journal}
  {\bibinfo  {journal} {Phys. Rev. B}\ }\textbf {\bibinfo {volume} {54}},\
  \bibinfo {pages} {11169} (\bibinfo {year} {1996}{\natexlab{a}})}\BibitemShut
  {NoStop}%
\bibitem [{\citenamefont {Muhli}\ and\ \citenamefont
  {Caro}(2024)}]{muhli_2024b}%
  \BibitemOpen
  \bibfield  {author} {\bibinfo {author} {\bibfnamefont {H.}~\bibnamefont
  {Muhli}}\ and\ \bibinfo {author} {\bibfnamefont {M.~A.}\ \bibnamefont
  {Caro}},\ }\bibfield  {title} {\enquote {\bibinfo {title} {General purpose
  {Gaussian} approximation potential for {P} with {Hirshfeld} volumes for {vdW}
  corrections},}\ }\href {\doibase DOI:10.5281/zenodo.14013797} {\bibfield
  {journal} {\bibinfo  {journal} {Zenodo}\ } (\bibinfo {year} {2024}),\
  DOI:10.5281/zenodo.14013797}\BibitemShut {NoStop}%
\bibitem [{\citenamefont {Larsen}\ \emph {et~al.}(2017)\citenamefont {Larsen},
  \citenamefont {Mortensen}, \citenamefont {Blomqvist}, \citenamefont
  {Castelli}, \citenamefont {Christensen}, \citenamefont {Dulak}, \citenamefont
  {Friis}, \citenamefont {Groves}, \citenamefont {Hammer}, \citenamefont
  {Hargus}, \citenamefont {Hermes}, \citenamefont {Jennings}, \citenamefont
  {Jensen}, \citenamefont {Kermode}, \citenamefont {Kitchin}, \citenamefont
  {Kolsbjerg}, \citenamefont {Kubal}, \citenamefont {Kaasbjerg}, \citenamefont
  {Lysgaard}, \citenamefont {Maronsson}, \citenamefont {Maxson}, \citenamefont
  {Olsen}, \citenamefont {Pastewka}, \citenamefont {Peterson}, \citenamefont
  {Rostgaard}, \citenamefont {Schi{\o}tz}, \citenamefont {Sch\"utt},
  \citenamefont {Strange}, \citenamefont {Thygesen}, \citenamefont {Vegge},
  \citenamefont {Vilhelmsen}, \citenamefont {Walter}, \citenamefont {Zeng},\
  and\ \citenamefont {Jacobsen}}]{larsen_2017}%
  \BibitemOpen
  \bibfield  {author} {\bibinfo {author} {\bibfnamefont {A.~H.}\ \bibnamefont
  {Larsen}}, \bibinfo {author} {\bibfnamefont {J.~J.}\ \bibnamefont
  {Mortensen}}, \bibinfo {author} {\bibfnamefont {J.}~\bibnamefont
  {Blomqvist}}, \bibinfo {author} {\bibfnamefont {I.~E.}\ \bibnamefont
  {Castelli}}, \bibinfo {author} {\bibfnamefont {R.}~\bibnamefont
  {Christensen}}, \bibinfo {author} {\bibfnamefont {M.}~\bibnamefont {Dulak}},
  \bibinfo {author} {\bibfnamefont {J.}~\bibnamefont {Friis}}, \bibinfo
  {author} {\bibfnamefont {M.~N.}\ \bibnamefont {Groves}}, \bibinfo {author}
  {\bibfnamefont {B.}~\bibnamefont {Hammer}}, \bibinfo {author} {\bibfnamefont
  {C.}~\bibnamefont {Hargus}}, \bibinfo {author} {\bibfnamefont {E.~D.}\
  \bibnamefont {Hermes}}, \bibinfo {author} {\bibfnamefont {P.~C.}\
  \bibnamefont {Jennings}}, \bibinfo {author} {\bibfnamefont {P.~B.}\
  \bibnamefont {Jensen}}, \bibinfo {author} {\bibfnamefont {J.}~\bibnamefont
  {Kermode}}, \bibinfo {author} {\bibfnamefont {J.~R.}\ \bibnamefont
  {Kitchin}}, \bibinfo {author} {\bibfnamefont {E.~L.}\ \bibnamefont
  {Kolsbjerg}}, \bibinfo {author} {\bibfnamefont {J.}~\bibnamefont {Kubal}},
  \bibinfo {author} {\bibfnamefont {K.}~\bibnamefont {Kaasbjerg}}, \bibinfo
  {author} {\bibfnamefont {S.}~\bibnamefont {Lysgaard}}, \bibinfo {author}
  {\bibfnamefont {J.~B.}\ \bibnamefont {Maronsson}}, \bibinfo {author}
  {\bibfnamefont {T.}~\bibnamefont {Maxson}}, \bibinfo {author} {\bibfnamefont
  {T.}~\bibnamefont {Olsen}}, \bibinfo {author} {\bibfnamefont
  {L.}~\bibnamefont {Pastewka}}, \bibinfo {author} {\bibfnamefont
  {A.}~\bibnamefont {Peterson}}, \bibinfo {author} {\bibfnamefont
  {C.}~\bibnamefont {Rostgaard}}, \bibinfo {author} {\bibfnamefont
  {J.}~\bibnamefont {Schi{\o}tz}}, \bibinfo {author} {\bibfnamefont
  {O.}~\bibnamefont {Sch\"utt}}, \bibinfo {author} {\bibfnamefont
  {M.}~\bibnamefont {Strange}}, \bibinfo {author} {\bibfnamefont {K.~S.}\
  \bibnamefont {Thygesen}}, \bibinfo {author} {\bibfnamefont {T.}~\bibnamefont
  {Vegge}}, \bibinfo {author} {\bibfnamefont {L.}~\bibnamefont {Vilhelmsen}},
  \bibinfo {author} {\bibfnamefont {M.}~\bibnamefont {Walter}}, \bibinfo
  {author} {\bibfnamefont {Z.}~\bibnamefont {Zeng}}, \ and\ \bibinfo {author}
  {\bibfnamefont {K.~W.}\ \bibnamefont {Jacobsen}},\ }\bibfield  {title}
  {\enquote {\bibinfo {title} {The {Atomic Simulation Environment} -- {A
  Python} library for working with atoms},}\ }\href@noop {} {\bibfield
  {journal} {\bibinfo  {journal} {J. Phys.: Condens. Matter}\ }\textbf
  {\bibinfo {volume} {29}},\ \bibinfo {pages} {273002} (\bibinfo {year}
  {2017})}\BibitemShut {NoStop}%
\bibitem [{\citenamefont {Stukowski}(2009)}]{stukowski_2009}%
  \BibitemOpen
  \bibfield  {author} {\bibinfo {author} {\bibfnamefont {A.}~\bibnamefont
  {Stukowski}},\ }\bibfield  {title} {\enquote {\bibinfo {title} {Visualization
  and analysis of atomistic simulation data with {OVITO--the Open Visualization
  Tool}},}\ }\href@noop {} {\bibfield  {journal} {\bibinfo  {journal}
  {Modelling Simul. Mater. Sci. Eng.}\ }\textbf {\bibinfo {volume} {18}},\
  \bibinfo {pages} {015012} (\bibinfo {year} {2009})}\BibitemShut {NoStop}%
\bibitem [{\citenamefont {Martin}(2008)}]{martin2008electronic}%
  \BibitemOpen
  \bibfield  {author} {\bibinfo {author} {\bibfnamefont {R.~M.}\ \bibnamefont
  {Martin}},\ }\href@noop {} {\emph {\bibinfo {title} {Electronic structure:
  basic theory and practical methods}}}\ (\bibinfo  {publisher} {Cambridge
  university press},\ \bibinfo {year} {2008})\BibitemShut {NoStop}%
\bibitem [{\citenamefont {St{\"o}hr}\ \emph {et~al.}(2019)\citenamefont
  {St{\"o}hr}, \citenamefont {Van~Voorhis},\ and\ \citenamefont
  {Tkatchenko}}]{stohr2019theory}%
  \BibitemOpen
  \bibfield  {author} {\bibinfo {author} {\bibfnamefont {M.}~\bibnamefont
  {St{\"o}hr}}, \bibinfo {author} {\bibfnamefont {T.}~\bibnamefont
  {Van~Voorhis}}, \ and\ \bibinfo {author} {\bibfnamefont {A.}~\bibnamefont
  {Tkatchenko}},\ }\bibfield  {title} {\enquote {\bibinfo {title} {Theory and
  practice of modeling van der {Waals} interactions in electronic-structure
  calculations},}\ }\href@noop {} {\bibfield  {journal} {\bibinfo  {journal}
  {Chem. Soc. Rev.}\ }\textbf {\bibinfo {volume} {48}},\ \bibinfo {pages}
  {4118} (\bibinfo {year} {2019})}\BibitemShut {NoStop}%
\bibitem [{\citenamefont {Dion}\ \emph {et~al.}(2004)\citenamefont {Dion},
  \citenamefont {Rydberg}, \citenamefont {Schr{\"o}der}, \citenamefont
  {Langreth},\ and\ \citenamefont {Lundqvist}}]{dion_2004}%
  \BibitemOpen
  \bibfield  {author} {\bibinfo {author} {\bibfnamefont {M.}~\bibnamefont
  {Dion}}, \bibinfo {author} {\bibfnamefont {H.}~\bibnamefont {Rydberg}},
  \bibinfo {author} {\bibfnamefont {E.}~\bibnamefont {Schr{\"o}der}}, \bibinfo
  {author} {\bibfnamefont {D.~C.}\ \bibnamefont {Langreth}}, \ and\ \bibinfo
  {author} {\bibfnamefont {B.~I.}\ \bibnamefont {Lundqvist}},\ }\bibfield
  {title} {\enquote {\bibinfo {title} {Van der {Waals} density functional for
  general geometries},}\ }\href@noop {} {\bibfield  {journal} {\bibinfo
  {journal} {Phys. Rev. Lett.}\ }\textbf {\bibinfo {volume} {92}},\ \bibinfo
  {pages} {246401} (\bibinfo {year} {2004})}\BibitemShut {NoStop}%
\bibitem [{\citenamefont {Klime\v{s}}\ \emph {et~al.}(2010)\citenamefont
  {Klime\v{s}}, \citenamefont {Bowler},\ and\ \citenamefont
  {Michaelides}}]{klimes_2010}%
  \BibitemOpen
  \bibfield  {author} {\bibinfo {author} {\bibfnamefont {J.}~\bibnamefont
  {Klime\v{s}}}, \bibinfo {author} {\bibfnamefont {D.~R.}\ \bibnamefont
  {Bowler}}, \ and\ \bibinfo {author} {\bibfnamefont {A.}~\bibnamefont
  {Michaelides}},\ }\bibfield  {title} {\enquote {\bibinfo {title} {Chemical
  accuracy for the {van der Waals} density functional},}\ }\href@noop {}
  {\bibfield  {journal} {\bibinfo  {journal} {J. Phys.: Condens. Matter}\
  }\textbf {\bibinfo {volume} {22}},\ \bibinfo {pages} {022201} (\bibinfo
  {year} {2010})}\BibitemShut {NoStop}%
\bibitem [{\citenamefont {Grimme}(2006)}]{grimme_2006}%
  \BibitemOpen
  \bibfield  {author} {\bibinfo {author} {\bibfnamefont {S.}~\bibnamefont
  {Grimme}},\ }\bibfield  {title} {\enquote {\bibinfo {title} {Semiempirical
  {GGA}-type density functional constructed with a long-range dispersion
  correction},}\ }\href@noop {} {\bibfield  {journal} {\bibinfo  {journal} {J.
  Comput. Chem.}\ }\textbf {\bibinfo {volume} {27}},\ \bibinfo {pages} {1787}
  (\bibinfo {year} {2006})}\BibitemShut {NoStop}%
\bibitem [{\citenamefont {Grimme}\ \emph {et~al.}(2010)\citenamefont {Grimme},
  \citenamefont {Antony}, \citenamefont {Ehrlich},\ and\ \citenamefont
  {Krieg}}]{grimme_2010}%
  \BibitemOpen
  \bibfield  {author} {\bibinfo {author} {\bibfnamefont {S.}~\bibnamefont
  {Grimme}}, \bibinfo {author} {\bibfnamefont {J.}~\bibnamefont {Antony}},
  \bibinfo {author} {\bibfnamefont {S.}~\bibnamefont {Ehrlich}}, \ and\
  \bibinfo {author} {\bibfnamefont {H.}~\bibnamefont {Krieg}},\ }\bibfield
  {title} {\enquote {\bibinfo {title} {A consistent and accurate ab initio
  parametrization of density functional dispersion correction (dft-d) for the
  94 elements h-pu},}\ }\href@noop {} {\bibfield  {journal} {\bibinfo
  {journal} {J. Chem. Phys.}\ }\textbf {\bibinfo {volume} {132}},\ \bibinfo
  {pages} {154104} (\bibinfo {year} {2010})}\BibitemShut {NoStop}%
\bibitem [{\citenamefont {Otero-de-la Roza}\ and\ \citenamefont
  {Johnson}(2013)}]{otero_2013}%
  \BibitemOpen
  \bibfield  {author} {\bibinfo {author} {\bibfnamefont {A.}~\bibnamefont
  {Otero-de-la Roza}}\ and\ \bibinfo {author} {\bibfnamefont {E.~R.}\
  \bibnamefont {Johnson}},\ }\bibfield  {title} {\enquote {\bibinfo {title}
  {Many-body dispersion interactions from the exchange-hole dipole moment
  model},}\ }\href@noop {} {\bibfield  {journal} {\bibinfo  {journal} {J. Chem.
  Phys.}\ }\textbf {\bibinfo {volume} {138}},\ \bibinfo {pages} {054103}
  (\bibinfo {year} {2013})}\BibitemShut {NoStop}%
\bibitem [{\citenamefont {Gunnarsson}\ \emph {et~al.}(1976)\citenamefont
  {Gunnarsson}, \citenamefont {Jonson},\ and\ \citenamefont
  {Lundqvist}}]{gunnarsson1976exchange}%
  \BibitemOpen
  \bibfield  {author} {\bibinfo {author} {\bibfnamefont {O.}~\bibnamefont
  {Gunnarsson}}, \bibinfo {author} {\bibfnamefont {M.}~\bibnamefont {Jonson}},
  \ and\ \bibinfo {author} {\bibfnamefont {B.}~\bibnamefont {Lundqvist}},\
  }\bibfield  {title} {\enquote {\bibinfo {title} {Exchange and correlation in
  atoms, molecules and clusters},}\ }\href@noop {} {\bibfield  {journal}
  {\bibinfo  {journal} {Phys. Lett. A}\ }\textbf {\bibinfo {volume} {59}},\
  \bibinfo {pages} {177} (\bibinfo {year} {1976})}\BibitemShut {NoStop}%
\bibitem [{\citenamefont {Gunnarsson}\ \emph {et~al.}(1977)\citenamefont
  {Gunnarsson}, \citenamefont {Jonson},\ and\ \citenamefont
  {Lundqvist}}]{gunnarsson1977exchange}%
  \BibitemOpen
  \bibfield  {author} {\bibinfo {author} {\bibfnamefont {O.}~\bibnamefont
  {Gunnarsson}}, \bibinfo {author} {\bibfnamefont {M.}~\bibnamefont {Jonson}},
  \ and\ \bibinfo {author} {\bibfnamefont {B.}~\bibnamefont {Lundqvist}},\
  }\bibfield  {title} {\enquote {\bibinfo {title} {Exchange and correlation in
  inhomogeneous electron systems},}\ }\href@noop {} {\bibfield  {journal}
  {\bibinfo  {journal} {Solid State Commun.}\ }\textbf {\bibinfo {volume}
  {24}},\ \bibinfo {pages} {765} (\bibinfo {year} {1977})}\BibitemShut
  {NoStop}%
\bibitem [{\citenamefont {Alonso}\ and\ \citenamefont
  {Girifalco}(1978)}]{alonso1978nonlocal}%
  \BibitemOpen
  \bibfield  {author} {\bibinfo {author} {\bibfnamefont {J.}~\bibnamefont
  {Alonso}}\ and\ \bibinfo {author} {\bibfnamefont {L.}~\bibnamefont
  {Girifalco}},\ }\bibfield  {title} {\enquote {\bibinfo {title} {Nonlocal
  approximation to the exchange potential and kinetic energy of an
  inhomogeneous electron gas},}\ }\href@noop {} {\bibfield  {journal} {\bibinfo
   {journal} {Phys. Rev. B}\ }\textbf {\bibinfo {volume} {17}},\ \bibinfo
  {pages} {3735} (\bibinfo {year} {1978})}\BibitemShut {NoStop}%
\bibitem [{\citenamefont {Gunnarsson}\ \emph {et~al.}(1979)\citenamefont
  {Gunnarsson}, \citenamefont {Jonson},\ and\ \citenamefont
  {Lundqvist}}]{gunnarsson1979descriptions}%
  \BibitemOpen
  \bibfield  {author} {\bibinfo {author} {\bibfnamefont {O.}~\bibnamefont
  {Gunnarsson}}, \bibinfo {author} {\bibfnamefont {M.}~\bibnamefont {Jonson}},
  \ and\ \bibinfo {author} {\bibfnamefont {B.}~\bibnamefont {Lundqvist}},\
  }\bibfield  {title} {\enquote {\bibinfo {title} {Descriptions of exchange and
  correlation effects in inhomogeneous electron systems},}\ }\href@noop {}
  {\bibfield  {journal} {\bibinfo  {journal} {Phys. Rev. B}\ }\textbf {\bibinfo
  {volume} {20}},\ \bibinfo {pages} {3136} (\bibinfo {year}
  {1979})}\BibitemShut {NoStop}%
\bibitem [{\citenamefont {Rom{\'a}n-P{\'e}rez}\ and\ \citenamefont
  {Soler}(2009)}]{roman2009efficient}%
  \BibitemOpen
  \bibfield  {author} {\bibinfo {author} {\bibfnamefont {G.}~\bibnamefont
  {Rom{\'a}n-P{\'e}rez}}\ and\ \bibinfo {author} {\bibfnamefont
  {J.}~\bibnamefont {Soler}},\ }\bibfield  {title} {\enquote {\bibinfo {title}
  {{Efficient Implementation of a van der Waals Density Functional: Application
  to Double-Wall Carbon Nanotubes}},}\ }\href@noop {} {\bibfield  {journal}
  {\bibinfo  {journal} {Phys. Rev. Lett.}\ }\textbf {\bibinfo {volume} {103}},\
  \bibinfo {pages} {096102} (\bibinfo {year} {2009})}\BibitemShut {NoStop}%
\bibitem [{\citenamefont {Klime{\v{s}}}\ \emph {et~al.}(2011)\citenamefont
  {Klime{\v{s}}}, \citenamefont {Bowler},\ and\ \citenamefont
  {Michaelides}}]{klimevs2011van}%
  \BibitemOpen
  \bibfield  {author} {\bibinfo {author} {\bibfnamefont {J.}~\bibnamefont
  {Klime{\v{s}}}}, \bibinfo {author} {\bibfnamefont {D.~R.}\ \bibnamefont
  {Bowler}}, \ and\ \bibinfo {author} {\bibfnamefont {A.}~\bibnamefont
  {Michaelides}},\ }\bibfield  {title} {\enquote {\bibinfo {title} {{Van der
  Waals density functionals applied to solids}},}\ }\href@noop {} {\bibfield
  {journal} {\bibinfo  {journal} {Phys. Rev. B}\ }\textbf {\bibinfo {volume}
  {83}},\ \bibinfo {pages} {195131} (\bibinfo {year} {2011})}\BibitemShut
  {NoStop}%
\bibitem [{\citenamefont {Hermann}\ and\ \citenamefont
  {Tkatchenko}(2020)}]{hermann2020density}%
  \BibitemOpen
  \bibfield  {author} {\bibinfo {author} {\bibfnamefont {J.}~\bibnamefont
  {Hermann}}\ and\ \bibinfo {author} {\bibfnamefont {A.}~\bibnamefont
  {Tkatchenko}},\ }\bibfield  {title} {\enquote {\bibinfo {title} {{Density
  functional model for van der Waals interactions: Unifying many-body atomic
  approaches with nonlocal functionals}},}\ }\href@noop {} {\bibfield
  {journal} {\bibinfo  {journal} {Phys. Rev. Lett.}\ }\textbf {\bibinfo
  {volume} {124}},\ \bibinfo {pages} {146401} (\bibinfo {year}
  {2020})}\BibitemShut {NoStop}%
\bibitem [{\citenamefont {Bereau}\ \emph {et~al.}(2018)\citenamefont {Bereau},
  \citenamefont {DiStasio~Jr}, \citenamefont {Tkatchenko},\ and\ \citenamefont
  {Von~Lilienfeld}}]{bereau_2018}%
  \BibitemOpen
  \bibfield  {author} {\bibinfo {author} {\bibfnamefont {T.}~\bibnamefont
  {Bereau}}, \bibinfo {author} {\bibfnamefont {R.~A.}\ \bibnamefont
  {DiStasio~Jr}}, \bibinfo {author} {\bibfnamefont {A.}~\bibnamefont
  {Tkatchenko}}, \ and\ \bibinfo {author} {\bibfnamefont {O.~A.}\ \bibnamefont
  {Von~Lilienfeld}},\ }\bibfield  {title} {\enquote {\bibinfo {title}
  {Non-covalent interactions across organic and biological subsets of chemical
  space: Physics-based potentials parametrized from machine learning},}\
  }\href@noop {} {\bibfield  {journal} {\bibinfo  {journal} {J. Chem. Phys.}\
  }\textbf {\bibinfo {volume} {148}},\ \bibinfo {pages} {241706} (\bibinfo
  {year} {2018})}\BibitemShut {NoStop}%
\bibitem [{\citenamefont {Ying}\ and\ \citenamefont {Fan}(2023)}]{ying_2023}%
  \BibitemOpen
  \bibfield  {author} {\bibinfo {author} {\bibfnamefont {P.}~\bibnamefont
  {Ying}}\ and\ \bibinfo {author} {\bibfnamefont {Z.}~\bibnamefont {Fan}},\
  }\bibfield  {title} {\enquote {\bibinfo {title} {Combining the {D3}
  dispersion correction with the neuroevolution machine-learned potential},}\
  }\href@noop {} {\bibfield  {journal} {\bibinfo  {journal} {J. Phys.: Condens.
  Matter}\ }\textbf {\bibinfo {volume} {36}},\ \bibinfo {pages} {125901}
  (\bibinfo {year} {2023})}\BibitemShut {NoStop}%
\bibitem [{\citenamefont {Tu}\ \emph {et~al.}(2024)\citenamefont {Tu},
  \citenamefont {Williamson}, \citenamefont {Johnson},\ and\ \citenamefont
  {Rowley}}]{tu_2024}%
  \BibitemOpen
  \bibfield  {author} {\bibinfo {author} {\bibfnamefont {N.~T.~P.}\
  \bibnamefont {Tu}}, \bibinfo {author} {\bibfnamefont {S.}~\bibnamefont
  {Williamson}}, \bibinfo {author} {\bibfnamefont {E.~R.}\ \bibnamefont
  {Johnson}}, \ and\ \bibinfo {author} {\bibfnamefont {C.~N.}\ \bibnamefont
  {Rowley}},\ }\bibfield  {title} {\enquote {\bibinfo {title} {Modeling
  intermolecular interactions with exchange-hole dipole moment dispersion
  corrections to neural network potentials},}\ }\href@noop {} {\bibfield
  {journal} {\bibinfo  {journal} {J. Phys. Chem. B}\ }\textbf {\bibinfo
  {volume} {128}},\ \bibinfo {pages} {8290} (\bibinfo {year}
  {2024})}\BibitemShut {NoStop}%
\bibitem [{\citenamefont {Hirshfeld}(1977)}]{hirshfeld1977bonded}%
  \BibitemOpen
  \bibfield  {author} {\bibinfo {author} {\bibfnamefont {F.}~\bibnamefont
  {Hirshfeld}},\ }\bibfield  {title} {\enquote {\bibinfo {title} {Bonded-atom
  fragments for describing molecular charge densities},}\ }\href@noop {}
  {\bibfield  {journal} {\bibinfo  {journal} {Theor. Chim. Acta}\ }\textbf
  {\bibinfo {volume} {44}},\ \bibinfo {pages} {129} (\bibinfo {year}
  {1977})}\BibitemShut {NoStop}%
\bibitem [{\citenamefont {Tkatchenko}\ \emph {et~al.}(2013)\citenamefont
  {Tkatchenko}, \citenamefont {Ambrosetti},\ and\ \citenamefont
  {DiStasio}}]{tkatchenko2013interatomic}%
  \BibitemOpen
  \bibfield  {author} {\bibinfo {author} {\bibfnamefont {A.}~\bibnamefont
  {Tkatchenko}}, \bibinfo {author} {\bibfnamefont {A.}~\bibnamefont
  {Ambrosetti}}, \ and\ \bibinfo {author} {\bibfnamefont {R.}~\bibnamefont
  {DiStasio}},\ }\bibfield  {title} {\enquote {\bibinfo {title} {Interatomic
  methods for the dispersion energy derived from the adiabatic connection
  fluctuation-dissipation theorem},}\ }\href@noop {} {\bibfield  {journal}
  {\bibinfo  {journal} {J. Chem. Phys.}\ }\textbf {\bibinfo {volume} {138}}
  (\bibinfo {year} {2013})}\BibitemShut {NoStop}%
\bibitem [{\citenamefont {Bu{\v{c}}ko}\ \emph {et~al.}(2016)\citenamefont
  {Bu{\v{c}}ko}, \citenamefont {Leb{\`e}gue}, \citenamefont {Gould},\ and\
  \citenamefont {{\'A}ngy{\'a}n}}]{buvcko2016many}%
  \BibitemOpen
  \bibfield  {author} {\bibinfo {author} {\bibfnamefont {T.}~\bibnamefont
  {Bu{\v{c}}ko}}, \bibinfo {author} {\bibfnamefont {S.}~\bibnamefont
  {Leb{\`e}gue}}, \bibinfo {author} {\bibfnamefont {T.}~\bibnamefont {Gould}},
  \ and\ \bibinfo {author} {\bibfnamefont {J.}~\bibnamefont {{\'A}ngy{\'a}n}},\
  }\bibfield  {title} {\enquote {\bibinfo {title} {Many-body dispersion
  corrections for periodic systems: an efficient reciprocal space
  implementation},}\ }\href@noop {} {\bibfield  {journal} {\bibinfo  {journal}
  {J. Phys.: Condens. Matter}\ }\textbf {\bibinfo {volume} {28}},\ \bibinfo
  {pages} {045201} (\bibinfo {year} {2016})}\BibitemShut {NoStop}%
\bibitem [{\citenamefont {Kresse}\ and\ \citenamefont
  {Hafner}(1993)}]{kresse1993ab}%
  \BibitemOpen
  \bibfield  {author} {\bibinfo {author} {\bibfnamefont {G.}~\bibnamefont
  {Kresse}}\ and\ \bibinfo {author} {\bibfnamefont {J.}~\bibnamefont
  {Hafner}},\ }\bibfield  {title} {\enquote {\bibinfo {title} {\textit{Ab
  initio} molecular dynamics for liquid metals},}\ }\href@noop {} {\bibfield
  {journal} {\bibinfo  {journal} {Phys. Rev. B}\ }\textbf {\bibinfo {volume}
  {47}},\ \bibinfo {pages} {558} (\bibinfo {year} {1993})}\BibitemShut
  {NoStop}%
\bibitem [{\citenamefont {Kresse}\ and\ \citenamefont
  {Furthm{\"u}ller}(1996{\natexlab{b}})}]{kresse1996efficiency}%
  \BibitemOpen
  \bibfield  {author} {\bibinfo {author} {\bibfnamefont {G.}~\bibnamefont
  {Kresse}}\ and\ \bibinfo {author} {\bibfnamefont {J.}~\bibnamefont
  {Furthm{\"u}ller}},\ }\bibfield  {title} {\enquote {\bibinfo {title}
  {Efficiency of ab-initio total energy calculations for metals and
  semiconductors using a plane-wave basis set},}\ }\href@noop {} {\bibfield
  {journal} {\bibinfo  {journal} {Comp. Mater. Sci.}\ }\textbf {\bibinfo
  {volume} {6}},\ \bibinfo {pages} {15} (\bibinfo {year}
  {1996}{\natexlab{b}})}\BibitemShut {NoStop}%
\bibitem [{\citenamefont {Thompson}\ \emph {et~al.}(2009)\citenamefont
  {Thompson}, \citenamefont {Plimpton},\ and\ \citenamefont
  {Mattson}}]{thompson2009general}%
  \BibitemOpen
  \bibfield  {author} {\bibinfo {author} {\bibfnamefont {A.}~\bibnamefont
  {Thompson}}, \bibinfo {author} {\bibfnamefont {S.}~\bibnamefont {Plimpton}},
  \ and\ \bibinfo {author} {\bibfnamefont {W.}~\bibnamefont {Mattson}},\
  }\bibfield  {title} {\enquote {\bibinfo {title} {General formulation of
  pressure and stress tensor for arbitrary many-body interaction potentials
  under periodic boundary conditions},}\ }\href@noop {} {\bibfield  {journal}
  {\bibinfo  {journal} {J. Chem. Phys.}\ }\textbf {\bibinfo {volume} {131}}
  (\bibinfo {year} {2009})}\BibitemShut {NoStop}%
\bibitem [{\citenamefont {Fan}\ \emph {et~al.}(2015)\citenamefont {Fan},
  \citenamefont {Pereira}, \citenamefont {Wang}, \citenamefont {Zheng},
  \citenamefont {Donadio},\ and\ \citenamefont {Harju}}]{fan2015force}%
  \BibitemOpen
  \bibfield  {author} {\bibinfo {author} {\bibfnamefont {Z.}~\bibnamefont
  {Fan}}, \bibinfo {author} {\bibfnamefont {L.}~\bibnamefont {Pereira}},
  \bibinfo {author} {\bibfnamefont {H.}~\bibnamefont {Wang}}, \bibinfo {author}
  {\bibfnamefont {J.}~\bibnamefont {Zheng}}, \bibinfo {author} {\bibfnamefont
  {D.}~\bibnamefont {Donadio}}, \ and\ \bibinfo {author} {\bibfnamefont
  {A.}~\bibnamefont {Harju}},\ }\bibfield  {title} {\enquote {\bibinfo {title}
  {Force and heat current formulas for many-body potentials in molecular
  dynamics simulations with applications to thermal conductivity
  calculations},}\ }\href@noop {} {\bibfield  {journal} {\bibinfo  {journal}
  {Phys. Rev. B}\ }\textbf {\bibinfo {volume} {92}},\ \bibinfo {pages} {094301}
  (\bibinfo {year} {2015})}\BibitemShut {NoStop}%
\bibitem [{\citenamefont {Subramaniyan}\ and\ \citenamefont
  {Sun}(2008)}]{subramaniyan2008continuum}%
  \BibitemOpen
  \bibfield  {author} {\bibinfo {author} {\bibfnamefont {A.}~\bibnamefont
  {Subramaniyan}}\ and\ \bibinfo {author} {\bibfnamefont {C.}~\bibnamefont
  {Sun}},\ }\bibfield  {title} {\enquote {\bibinfo {title} {Continuum
  interpretation of virial stress in molecular simulations},}\ }\href@noop {}
  {\bibfield  {journal} {\bibinfo  {journal} {Int. J. Solids Struct.}\ }\textbf
  {\bibinfo {volume} {45}},\ \bibinfo {pages} {4340} (\bibinfo {year}
  {2008})}\BibitemShut {NoStop}%
\bibitem [{\citenamefont {Gutowsky}\ and\ \citenamefont
  {Hoffman}(1950)}]{gutowsky_1950}%
  \BibitemOpen
  \bibfield  {author} {\bibinfo {author} {\bibfnamefont {H.~S.}\ \bibnamefont
  {Gutowsky}}\ and\ \bibinfo {author} {\bibfnamefont {C.~J.}\ \bibnamefont
  {Hoffman}},\ }\bibfield  {title} {\enquote {\bibinfo {title} {The infrared
  spectrum of $p_4$},}\ }\href@noop {} {\bibfield  {journal} {\bibinfo
  {journal} {J. Am. Chem. Soc.}\ }\textbf {\bibinfo {volume} {72}},\ \bibinfo
  {pages} {5751} (\bibinfo {year} {1950})}\BibitemShut {NoStop}%
\bibitem [{\citenamefont {Chikvaidze}\ and\ \citenamefont
  {Gabrusenoks}(2023)}]{chikvaidze_2023}%
  \BibitemOpen
  \bibfield  {author} {\bibinfo {author} {\bibfnamefont {G.}~\bibnamefont
  {Chikvaidze}}\ and\ \bibinfo {author} {\bibfnamefont {J.}~\bibnamefont
  {Gabrusenoks}},\ }\bibfield  {title} {\enquote {\bibinfo {title} {Vibrational
  spectra and lattice dynamics of the $\beta$-phase of white phosphorus},}\
  }\href@noop {} {\bibfield  {journal} {\bibinfo  {journal} {J. Phys. Chem.
  Solids}\ }\textbf {\bibinfo {volume} {178}},\ \bibinfo {pages} {111356}
  (\bibinfo {year} {2023})}\BibitemShut {NoStop}%
\bibitem [{\citenamefont {Streett}\ \emph {et~al.}(1978)\citenamefont
  {Streett}, \citenamefont {Tildesley},\ and\ \citenamefont
  {Saville}}]{streett_1978}%
  \BibitemOpen
  \bibfield  {author} {\bibinfo {author} {\bibfnamefont {W.~B.}\ \bibnamefont
  {Streett}}, \bibinfo {author} {\bibfnamefont {D.~J.}\ \bibnamefont
  {Tildesley}}, \ and\ \bibinfo {author} {\bibfnamefont {G.}~\bibnamefont
  {Saville}},\ }\bibfield  {title} {\enquote {\bibinfo {title} {Multiple
  time-step methods in molecular dynamics},}\ }\href@noop {} {\bibfield
  {journal} {\bibinfo  {journal} {Mol. Phys.}\ }\textbf {\bibinfo {volume}
  {35}},\ \bibinfo {pages} {639} (\bibinfo {year} {1978})}\BibitemShut
  {NoStop}%
\bibitem [{\citenamefont {Miller}\ and\ \citenamefont
  {Bederson}(1978)}]{miller1978atomic}%
  \BibitemOpen
  \bibfield  {author} {\bibinfo {author} {\bibfnamefont {T.}~\bibnamefont
  {Miller}}\ and\ \bibinfo {author} {\bibfnamefont {B.}~\bibnamefont
  {Bederson}},\ }\bibfield  {title} {\enquote {\bibinfo {title} {{Atomic and
  Molecular Polarizabilities-A Review of Recent Advances}},}\ }\href@noop {}
  {\bibfield  {journal} {\bibinfo  {journal} {Adv. Atom. Mol. Phys.}\ }\textbf
  {\bibinfo {volume} {13}},\ \bibinfo {pages} {1} (\bibinfo {year}
  {1978})}\BibitemShut {NoStop}%
\bibitem [{\citenamefont {Ballard}\ \emph {et~al.}(2000)\citenamefont
  {Ballard}, \citenamefont {Bonin},\ and\ \citenamefont
  {Louderback}}]{ballard2000absolute}%
  \BibitemOpen
  \bibfield  {author} {\bibinfo {author} {\bibfnamefont {A.}~\bibnamefont
  {Ballard}}, \bibinfo {author} {\bibfnamefont {K.}~\bibnamefont {Bonin}}, \
  and\ \bibinfo {author} {\bibfnamefont {J.}~\bibnamefont {Louderback}},\
  }\bibfield  {title} {\enquote {\bibinfo {title} {{Absolute measurement of the
  optical polarizability of C$_{60}$}},}\ }\href@noop {} {\bibfield  {journal}
  {\bibinfo  {journal} {J. Chem. Phys.}\ }\textbf {\bibinfo {volume} {113}},\
  \bibinfo {pages} {5732} (\bibinfo {year} {2000})}\BibitemShut {NoStop}%
\bibitem [{\citenamefont {Thakkar}\ and\ \citenamefont
  {Wu}(2015)}]{thakkar2015well}%
  \BibitemOpen
  \bibfield  {author} {\bibinfo {author} {\bibfnamefont {A.}~\bibnamefont
  {Thakkar}}\ and\ \bibinfo {author} {\bibfnamefont {T.}~\bibnamefont {Wu}},\
  }\bibfield  {title} {\enquote {\bibinfo {title} {{How well do static
  electronic dipole polarizabilities from gas-phase experiments compare with
  density functional and MP2 computations?}}}\ }\href@noop {} {\bibfield
  {journal} {\bibinfo  {journal} {J. Chem. Phys.}\ }\textbf {\bibinfo {volume}
  {143}},\ \bibinfo {pages} {144302} (\bibinfo {year} {2015})}\BibitemShut
  {NoStop}%
\bibitem [{\citenamefont {Bu\v{c}ko}\ \emph {et~al.}(2013)\citenamefont
  {Bu\v{c}ko}, \citenamefont {Leb\`egue}, \citenamefont {Hafner},\ and\
  \citenamefont {\'Angy\'an}}]{bucko_2013}%
  \BibitemOpen
  \bibfield  {author} {\bibinfo {author} {\bibfnamefont {T.}~\bibnamefont
  {Bu\v{c}ko}}, \bibinfo {author} {\bibfnamefont {S.}~\bibnamefont
  {Leb\`egue}}, \bibinfo {author} {\bibfnamefont {J.}~\bibnamefont {Hafner}}, \
  and\ \bibinfo {author} {\bibfnamefont {J.~G.}\ \bibnamefont {\'Angy\'an}},\
  }\bibfield  {title} {\enquote {\bibinfo {title} {Improved density dependent
  correction for the description of {London} dispersion forces},}\ }\href@noop
  {} {\bibfield  {journal} {\bibinfo  {journal} {J. Chem. Theory Comput.}\
  }\textbf {\bibinfo {volume} {9}},\ \bibinfo {pages} {4293} (\bibinfo {year}
  {2013})}\BibitemShut {NoStop}%
\bibitem [{\citenamefont {Gould}\ \emph {et~al.}(2016)\citenamefont {Gould},
  \citenamefont {Leb\`egue}, \citenamefont {\'Angy\'an},\ and\ \citenamefont
  {Bu\v{c}ko}}]{gould_2016}%
  \BibitemOpen
  \bibfield  {author} {\bibinfo {author} {\bibfnamefont {T.}~\bibnamefont
  {Gould}}, \bibinfo {author} {\bibfnamefont {S.}~\bibnamefont {Leb\`egue}},
  \bibinfo {author} {\bibfnamefont {J.~G.}\ \bibnamefont {\'Angy\'an}}, \ and\
  \bibinfo {author} {\bibfnamefont {T.}~\bibnamefont {Bu\v{c}ko}},\ }\bibfield
  {title} {\enquote {\bibinfo {title} {A fractionally ionic approach to
  polarizability and van der {Waals} many-body dispersion calculations},}\
  }\href@noop {} {\bibfield  {journal} {\bibinfo  {journal} {J. Chem. Theory
  Comput.}\ }\textbf {\bibinfo {volume} {12}},\ \bibinfo {pages} {5920}
  (\bibinfo {year} {2016})}\BibitemShut {NoStop}%
\bibitem [{\citenamefont {Bussi}\ \emph {et~al.}(2007)\citenamefont {Bussi},
  \citenamefont {Donadio},\ and\ \citenamefont {Parrinello}}]{bussi_2007}%
  \BibitemOpen
  \bibfield  {author} {\bibinfo {author} {\bibfnamefont {G.}~\bibnamefont
  {Bussi}}, \bibinfo {author} {\bibfnamefont {D.}~\bibnamefont {Donadio}}, \
  and\ \bibinfo {author} {\bibfnamefont {M.}~\bibnamefont {Parrinello}},\
  }\bibfield  {title} {\enquote {\bibinfo {title} {Canonical sampling through
  velocity rescaling},}\ }\href@noop {} {\bibfield  {journal} {\bibinfo
  {journal} {J. Chem. Phys.}\ }\textbf {\bibinfo {volume} {126}},\ \bibinfo
  {pages} {014101} (\bibinfo {year} {2007})}\BibitemShut {NoStop}%
\bibitem [{\citenamefont {Berendsen}\ \emph {et~al.}(1984)\citenamefont
  {Berendsen}, \citenamefont {Postma}, \citenamefont {van Gunsteren},
  \citenamefont {DiNola},\ and\ \citenamefont {Haak}}]{berendsen_1984}%
  \BibitemOpen
  \bibfield  {author} {\bibinfo {author} {\bibfnamefont {H.~J.~C.}\
  \bibnamefont {Berendsen}}, \bibinfo {author} {\bibfnamefont {J.~P.~M.}\
  \bibnamefont {Postma}}, \bibinfo {author} {\bibfnamefont {W.~F.}\
  \bibnamefont {van Gunsteren}}, \bibinfo {author} {\bibfnamefont
  {A.~R.~H.~J.}\ \bibnamefont {DiNola}}, \ and\ \bibinfo {author}
  {\bibfnamefont {J.~R.}\ \bibnamefont {Haak}},\ }\bibfield  {title} {\enquote
  {\bibinfo {title} {Molecular dynamics with coupling to an external bath},}\
  }\href@noop {} {\bibfield  {journal} {\bibinfo  {journal} {J. Chem. Phys.}\
  }\textbf {\bibinfo {volume} {81}},\ \bibinfo {pages} {3684} (\bibinfo {year}
  {1984})}\BibitemShut {NoStop}%
\bibitem [{\citenamefont {Caro}\ \emph {et~al.}(2016)\citenamefont {Caro},
  \citenamefont {Laurila},\ and\ \citenamefont {Lopez-Acevedo}}]{caro_2016}%
  \BibitemOpen
  \bibfield  {author} {\bibinfo {author} {\bibfnamefont {M.~A.}\ \bibnamefont
  {Caro}}, \bibinfo {author} {\bibfnamefont {T.}~\bibnamefont {Laurila}}, \
  and\ \bibinfo {author} {\bibfnamefont {O.}~\bibnamefont {Lopez-Acevedo}},\
  }\bibfield  {title} {\enquote {\bibinfo {title} {Accurate schemes for
  calculation of thermodynamic properties of liquid mixtures from molecular
  dynamics simulations},}\ }\href@noop {} {\bibfield  {journal} {\bibinfo
  {journal} {J. Chem. Phys.}\ }\textbf {\bibinfo {volume} {145}},\ \bibinfo
  {pages} {244504} (\bibinfo {year} {2016})}\BibitemShut {NoStop}%
\bibitem [{\citenamefont {Caro}\ \emph {et~al.}(2017)\citenamefont {Caro},
  \citenamefont {Lopez-Acevedo},\ and\ \citenamefont {Laurila}}]{caro_2017b}%
  \BibitemOpen
  \bibfield  {author} {\bibinfo {author} {\bibfnamefont {M.~A.}\ \bibnamefont
  {Caro}}, \bibinfo {author} {\bibfnamefont {O.}~\bibnamefont {Lopez-Acevedo}},
  \ and\ \bibinfo {author} {\bibfnamefont {T.}~\bibnamefont {Laurila}},\
  }\bibfield  {title} {\enquote {\bibinfo {title} {Redox potentials from ab
  initio molecular dynamics and explicit entropy calculations: application to
  transition metals in aqueous solution},}\ }\href@noop {} {\bibfield
  {journal} {\bibinfo  {journal} {J. Chem. Theory Comput.}\ }\textbf {\bibinfo
  {volume} {13}},\ \bibinfo {pages} {3432} (\bibinfo {year}
  {2017})}\BibitemShut {NoStop}%
\bibitem [{ref(last accessed: 2024-10-30)}]{ref_dospt}%
  \BibitemOpen
  \href@noop {} {\enquote {\bibinfo {title} {{DoSPT code}},}\ }\bibinfo
  {howpublished} {\url{http://dospt.org}} (\bibinfo {year} {last accessed:
  2024-10-30})\BibitemShut {NoStop}%
\bibitem [{\citenamefont {Lin}\ \emph {et~al.}(2003)\citenamefont {Lin},
  \citenamefont {Blanco},\ and\ \citenamefont {Goddard~{III}}}]{lin_2003}%
  \BibitemOpen
  \bibfield  {author} {\bibinfo {author} {\bibfnamefont {S.-T.}\ \bibnamefont
  {Lin}}, \bibinfo {author} {\bibfnamefont {M.}~\bibnamefont {Blanco}}, \ and\
  \bibinfo {author} {\bibfnamefont {W.~A.}\ \bibnamefont {Goddard~{III}}},\
  }\bibfield  {title} {\enquote {\bibinfo {title} {The two-phase model for
  calculating thermodynamic properties of liquids from molecular dynamics:
  Validation for the phase diagram of {Lennard-Jones} fluids},}\ }\href@noop {}
  {\bibfield  {journal} {\bibinfo  {journal} {J. Chem. Phys.}\ }\textbf
  {\bibinfo {volume} {119}},\ \bibinfo {pages} {11792} (\bibinfo {year}
  {2003})}\BibitemShut {NoStop}%
\bibitem [{\citenamefont {Bridgman}(1914)}]{bridgman_1914}%
  \BibitemOpen
  \bibfield  {author} {\bibinfo {author} {\bibfnamefont {P.~W.}\ \bibnamefont
  {Bridgman}},\ }\bibfield  {title} {\enquote {\bibinfo {title} {Two new
  modifications of phosphorus.}}\ }\href@noop {} {\bibfield  {journal}
  {\bibinfo  {journal} {J. Am. Chem. Soc.}\ }\textbf {\bibinfo {volume} {36}},\
  \bibinfo {pages} {1344} (\bibinfo {year} {1914})}\BibitemShut {NoStop}%
\bibitem [{\citenamefont {Morris}\ and\ \citenamefont
  {Song}(2002)}]{morris_2002}%
  \BibitemOpen
  \bibfield  {author} {\bibinfo {author} {\bibfnamefont {J.~R.}\ \bibnamefont
  {Morris}}\ and\ \bibinfo {author} {\bibfnamefont {X.}~\bibnamefont {Song}},\
  }\bibfield  {title} {\enquote {\bibinfo {title} {The melting lines of model
  systems calculated from coexistence simulations},}\ }\href@noop {} {\bibfield
   {journal} {\bibinfo  {journal} {J. Chem. Phys.}\ }\textbf {\bibinfo {volume}
  {116}},\ \bibinfo {pages} {9352} (\bibinfo {year} {2002})}\BibitemShut
  {NoStop}%
\bibitem [{\citenamefont {P\'artay}\ \emph {et~al.}(2021)\citenamefont
  {P\'artay}, \citenamefont {Cs\'anyi},\ and\ \citenamefont
  {Bernstein}}]{partay_2021}%
  \BibitemOpen
  \bibfield  {author} {\bibinfo {author} {\bibfnamefont {L.~B.}\ \bibnamefont
  {P\'artay}}, \bibinfo {author} {\bibfnamefont {G.}~\bibnamefont {Cs\'anyi}},
  \ and\ \bibinfo {author} {\bibfnamefont {N.}~\bibnamefont {Bernstein}},\
  }\bibfield  {title} {\enquote {\bibinfo {title} {Nested sampling for
  materials},}\ }\href@noop {} {\bibfield  {journal} {\bibinfo  {journal} {Eur.
  Phys. J. B}\ }\textbf {\bibinfo {volume} {94}},\ \bibinfo {pages} {159}
  (\bibinfo {year} {2021})}\BibitemShut {NoStop}%
\bibitem [{\citenamefont {Jana}\ and\ \citenamefont {Caro}(2023)}]{jana_2023}%
  \BibitemOpen
  \bibfield  {author} {\bibinfo {author} {\bibfnamefont {R.}~\bibnamefont
  {Jana}}\ and\ \bibinfo {author} {\bibfnamefont {M.~A.}\ \bibnamefont
  {Caro}},\ }\bibfield  {title} {\enquote {\bibinfo {title} {Searching for iron
  nanoparticles with a general-purpose {Gaussian} approximation potential},}\
  }\href@noop {} {\bibfield  {journal} {\bibinfo  {journal} {Phys. Rev. B}\
  }\textbf {\bibinfo {volume} {107}},\ \bibinfo {pages} {245421} (\bibinfo
  {year} {2023})}\BibitemShut {NoStop}%
\bibitem [{\citenamefont {Kloppenburg}\ \emph {et~al.}(2023)\citenamefont
  {Kloppenburg}, \citenamefont {P{\'a}rtay}, \citenamefont {J{\'o}nsson},\ and\
  \citenamefont {Caro}}]{kloppenburg_2023}%
  \BibitemOpen
  \bibfield  {author} {\bibinfo {author} {\bibfnamefont {J.}~\bibnamefont
  {Kloppenburg}}, \bibinfo {author} {\bibfnamefont {L.~B.}\ \bibnamefont
  {P{\'a}rtay}}, \bibinfo {author} {\bibfnamefont {H.}~\bibnamefont
  {J{\'o}nsson}}, \ and\ \bibinfo {author} {\bibfnamefont {M.~A.}\ \bibnamefont
  {Caro}},\ }\bibfield  {title} {\enquote {\bibinfo {title} {{A general-purpose
  machine learning Pt interatomic potential for an accurate description of
  bulk, surfaces, and nanoparticles}},}\ }\href@noop {} {\bibfield  {journal}
  {\bibinfo  {journal} {J. Chem. Phys.}\ }\textbf {\bibinfo {volume} {158}},\
  \bibinfo {pages} {134704} (\bibinfo {year} {2023})}\BibitemShut {NoStop}%
\end{thebibliography}
\end{document}